\documentclass[journal]{IEEEtran}
\usepackage{lineno}
\usepackage[normalem]{ulem}
\usepackage{physics}
\usepackage{epsfig}
\usepackage{breqn}
\usepackage{multirow}
\usepackage{amsmath}
\usepackage{xcolor}
\usepackage{verbatim}
\usepackage{tabularx}
\usepackage{tabulary}
\usepackage{tabulary}
\usepackage{booktabs}
\usepackage[utf8]{inputenc}
\UseRawInputEncoding
\usepackage[T1]{fontenc}
\usepackage[utf8]{inputenc}  % only for pdflatex
\usepackage{amsmath, amssymb, amsfonts}
\ifCLASSINFOpdf
  
\else
 
\fi

\begin{document}

\title{Fiber Bragg Grating Sensors with Enhanced Sensitivity for High-Precision Multi-Parameter Detection}

\author{Neha~Ahlawat         and~Saurabh~Mani~Tripathi,~\IEEEmembership{Senior~Member,~IEEE, Senior~Member,~Optica}% <-this % stops a space
\thanks{Authors are with Optics and Photonics Centre, Indian Institute of Technology Delhi, New Delhi, 110016, India. E-mail: N. Ahlawat (neha.ahlawat@iitd.ac.in), S. M. Triapthi (smt@iitd.ac.in).}% <-this % stops a space
\thanks{Manuscript received April 19, 2005; revised August 26, 2015.}}

\markboth{Journal of \LaTeX\ Class Files,~Vol.~14, No.~8, August~2015}%
{Shell \MakeLowercase{\textit{et al.}}: Bare Demo of IEEEtran.cls for IEEE Journals}

\maketitle

\begin{abstract}
Exceptional points (EPs), intrinsic to non-Hermitian systems, exhibit singular spectral responses with extreme sensitivity to external perturbations, offering new opportunities for precision sensing. In this work, we investigate the sensing performance of Fiber Bragg Gratings (FBGs) engineered to operate near EPs through precise structural tuning. By aligning the reflection spectrum edges with the EP condition, significant sensitivity enhancement is achieved under a power interrogation scheme. The optimized sensors demonstrate sensitivities of $\sim$9.3 dBm/ $^{\circ}$C for temperature and 0.5 dBm/$\mu\epsilon$ for axial strain, representing nearly an order-of-magnitude improvement over conventional FBG sensors. %, 2.9 dBm/$\mu\epsilon$, and 0.2 dBm/$\mu\epsilon$, corresponding to improvements of nearly an order of magnitude over conventional FBG sensors for temperature, axial strain, radial strain, and hydraulic strain perturbations, respectively. 
To address cross-sensitivity between temperature and strains, a sensitivity matrix-based approach is implemented, enabling accurate simultaneous detection of multiple perturbations. These results establish EP-engineered FBGs as a highly effective and versatile platform for high-precision, multi-parameter sensing, with broad applicability in telecommunications, structural health monitoring, aerospace systems, and environmental diagnostics.
\end{abstract}


\IEEEpeerreviewmaketitle

\section{Introduction}

\IEEEPARstart{F}{iber} Bragg Gratings (FBGs), consisting of periodic refractive index modulations inscribed within the core of optical fibers, are foundational elements in modern photonic technologies. Their unique ability to reflect specific wavelengths, where phase-matching conditions are satisfied, while transmitting others has made them indispensable in optical communication networks, signal processing systems, and a broad range of sensing applications \cite{kashyap_book, othonos_fbg, hill_fbg}. In particular, FBGs have been widely deployed in wavelength-division multiplexing (WDM) systems for telecommunications, structural health monitoring in civil and aerospace engineering, and biomedical diagnostics for detecting subtle physiological variations \cite{othonos_fbg, hill_fbg, Erdogan_fbg_review}.

Traditionally, research on FBGs has focused on optimizing their linear optical properties, reflectivity, bandwidth, and spectral response, which directly influence sensor resolution and device performance \cite{Erdogan_fbg_review, kashyap_book}. With the advancement of fabrication techniques, including phase mask lithography and femtosecond laser inscription, highly controlled grating profiles with enhanced performance have been realized \cite{kashyap_book, Mihailov}. The sensing mechanism of FBGs primarily relies on monitoring shifts in the Bragg wavelength, which directly respond to changes in environmental parameters such as temperature and strain. In standard silica fibers, temperature and strain sensitivities are typically $\sim10 pm/^{\circ}$C and $\sim1.2 pm/\mu\epsilon$, respectively \cite{hill_fbg}. Several recent studies have reported improved sensitivities, including temperature sensitivities up to 18.4 pm/$^{\circ}$C at longer wavelengths \cite{Gunday}, neural-network-enhanced designs achieving 0.8625 pm/$^{\circ}$C sensitivity \cite{Zahraa}, and strain sensitivities reaching 6.2 pm/$\mu\epsilon$ \cite{Li}. Further advances employing photonic crystal fibers, functional coatings, and composite integration have led to improved sensitivity, mechanical robustness, and long-term stability \cite{Zhou_2022,Zhang_2021,Wang_2022, Li_2024}.

A persistent challenge, however, remains the inherent cross-sensitivity between temperature and strain, which complicates accurate multi-parameter measurements. Various strategies have been explored to mitigate this issue, including dual-grating configurations, tailored coatings, hybrid structures, and advanced calibration protocols \cite{Qin}. Coating materials such as polymers or metals can enhance temperature sensitivity while partially decoupling strain effects \cite{Ganainy}. More recently, artificial intelligence (AI)-driven signal processing techniques have shown promise in effectively disentangling temperature and strain contributions, enabling more accurate and robust sensing under complex operational conditions \cite{ Özdemir}.

Beyond these conventional approaches, recent developments in non-Hermitian photonics have introduced fundamentally new paradigms for enhancing FBG functionality. In particular, the concept of exceptional points (EPs), non-Hermitian degeneracies where both eigenvalues and eigenvectors coalesce, has opened new avenues for designing ultra-sensitive photonic systems \cite{Heiss, Miri_EP_review,Chen_EP_sensing}. Near EPs, systems exhibit nontrivial spectral behavior, including extreme sensitivity to external perturbations, nonreciprocal light transmission, and enhanced nonlinear effects, phenomena absent in conventional Hermitian systems \cite{Hodaei_EP_sensing, Wiersig_EP_sensing,G652}.  %\textcolor{blue}{\uline{Kamyar Behrouzi et al. presented an integrated plasmonic EP sensor using two distinct resonators, demonstrating coalescence of mode frequency and loss rates, enabling ultra-sensitive detection down to single-molecule levels \cite{Behro}.}}
By leveraging EP physics, FBGs can be engineered to exhibit dramatically amplified responses to minute environmental changes, making them highly attractive for precision sensing of temperature, pressure, and multiple forms of strain.

In this work, we present a systematic study on the integration of exceptional point physics into FBG-based sensing platforms. By carefully engineering the grating profile, we achieve spectral coalescence between the FBG band edges and the exceptional points, resulting in substantial sensitivity enhancement. Specifically, we demonstrate sensitivity improvements for multiple perturbation parameters, including temperature, axial strain, radial strain, and hydraulic strain. Furthermore, we implement a sensitivity matrix-based approach that enables accurate simultaneous measurement of temperature and multiple strain components, effectively addressing the long-standing cross-sensitivity challenge in FBG sensors. The proposed EP-engineered FBG platform offers a versatile and powerful solution for high-precision, multi-parameter sensing, with broad relevance across telecommunications, structural monitoring, aerospace systems, and biomedical diagnostics.

Unlike PT-symmetric platforms which rely on balancing gain and loss, the present FBG system realizes EPs through passive coupling and distributed scattering loss. The resulting EPs share the same square root singularity characteristics as those in Smatrix formulations. However, they differ in their physical origin, a detailed comparison with \cite{science.abj3179} highlights that our EPs are inherently lossy rather than gain-loss balanced. Furthermore, recent advances in EP-enhanced fiber devices \cite{Behrouzi} reinforce the novelty of our approach.

\section{Theoretical Analysis}
\begin{figure}
    \centering
    \includegraphics[width=0.95\linewidth]{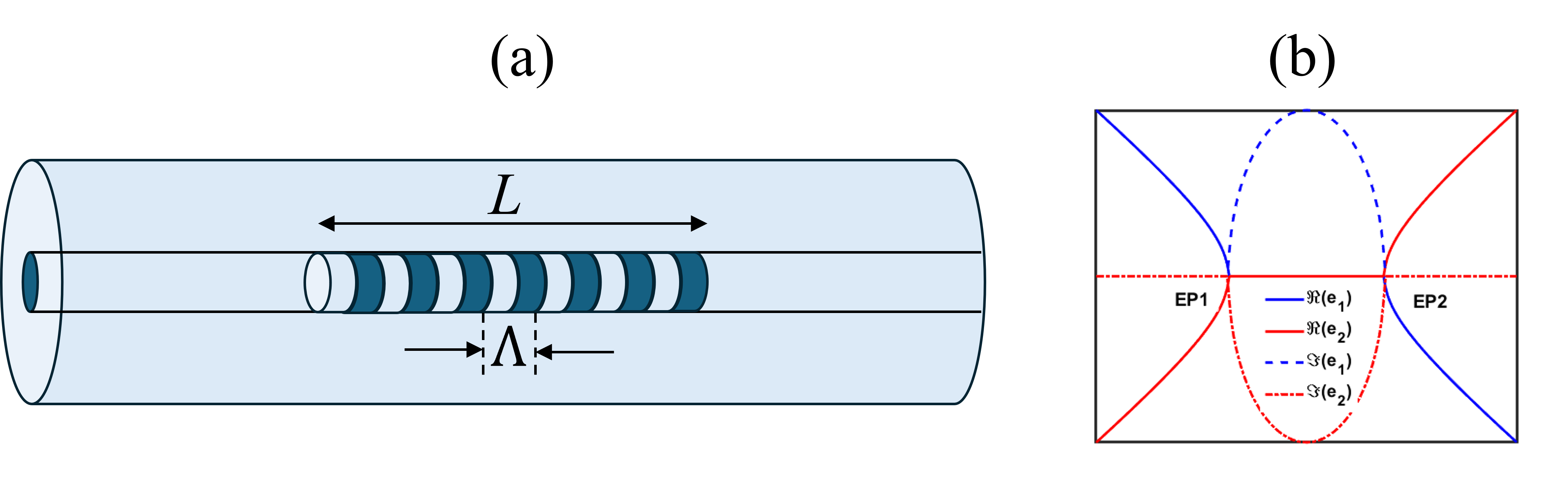}
    \caption{Schematic diagram of (a) the FBG and (b) EP at coalescing points of $\Re$ and $\Im$ parts of the eigenvalues.}
    \label{schematic}
\end{figure}

The schematic diagram of the sensor and EPs, generated at the coalescing points of $\Re$ and $\Im$ parts of the eigenvalues, are symbolically shown in Fig.~\ref{schematic}(a) and Fig.~\ref{schematic}(b), respectively.

The mode coupling within the FBGs is analytically described by coupled-mode equations that govern the forward and backward propagating fields, $A(z)$ and $B(z)$, respectively, through \cite{kashyap_book}:
\begin{equation*}
\begin{aligned}
\frac{dA(z)}{dz} &= -i\delta A(z) - i\kappa_{ac} B(z), \\
\frac{dB(z)}{dz} &= \;\; i\delta B(z) + i\kappa_{ac} A(z).
\end{aligned}
\end{equation*}

Here $\delta = \frac{1}{2}(\beta_A + \beta_B - \frac{2 \pi}{\Lambda})+ \kappa_{dc A} + \kappa_{dc B}$, represents the detuning, $\kappa_{dc A/B}$ is self-coupling (dc-coupling) coefficient, $\kappa_{ac}$ is the cross-coupling coefficient, and $z$ is the position direction. 
\\
Rewriting the coupled equations in matrix form:
\begin{center}
 \begin{math}
\frac{d}{dz} \begin{bmatrix} A(z) \\ B(z) \end{bmatrix} = \begin{bmatrix}
-i\delta & -i\kappa_{ac} \\
i\kappa_{ac} & i\delta 
\end{bmatrix} \begin{bmatrix} A(z) \\ B(z) \end{bmatrix} 
\end{math}
   
\end{center}
\begin{equation}
\Rightarrow
i\frac{d}{dz}
\begin{bmatrix}
    R \\
    S
\end{bmatrix}
=
\hat{H}
\begin{bmatrix}
    R \\
    S
\end{bmatrix},
\text{ with } 
\hat{H} =
\begin{bmatrix}
  \delta && \kappa_{ac}^*\\ -\kappa_{ac} && -\delta 
\end{bmatrix}.
\end{equation}
These coupled mode equations share a striking mathematical resemblance with the time dependent Schrödinger equation, with the only difference being that the role of time $t$ in quantum mechanics is replaced by the spatial coordinate 
$z$ along the fiber. 
The coupling coefficient in the FBG system acts similarly to the potential term, influencing the mode amplitudes as they propagate through the grating. This analogy allows us to draw insights from quantum mechanics, such as tunneling and reflection phenomena, to understand light propagation in FBGs.

Diagonalizing the Hamiltonian $\hat{H}$, as defined in Eq.(1), gives the eigenvalues as $e_{1,2} = \pm \sqrt{|\kappa_{ac}|^2-\delta^2}$. The corresponding eigenfunctions are $| \psi_1 \rangle = -i \frac{\delta+\sqrt{|\kappa_{ac}|^2-\delta^2}}{\kappa_{ac}}$ and $| \psi_2 \rangle = i \frac{-\delta+\sqrt{|\kappa_{ac}|^2-\delta^2}}{\kappa_{ac}}$, respectively. 
%
%We observe that at \(|\kappa_{ac}| = \pm \delta\), the two eigenvalues, as well as the corresponding eigenfunctions, coalesce, signifying the presence of exceptional points (EPs) when this condition is met. These points mark a transition where the system's eigenstates become indistinguishable, leading to non-Hermitian degeneracies. 
%
%This square-root dependence is the hallmark of an exceptional point (EP): at $|\delta|=|\kappa_{ac}|$, both eigenvalues and eigenvectors coalesce.
%
%Interestingly, \(|\kappa_{ac}| = \pm \delta\) also corresponds to the band-edge of the reflection spectrum in FBGs. In conventional Hermitian systems, band-edges define the transition between high- and low- transmission regimes. However, the band-edges are often not very sharp, and the transition from high- to low- reflectivity is often very smooth. 
%In the presence of EPs, the spectral response undergoes a dramatic transformation because of the square-root-like dispersion relation near these points. 

\subsection{Topological characteristics of the Exceptional Point in the Coupled-Mode FBG System}
The square-root functional form encodes the presence of a branch-point singularity: when $|\delta|=|\kappa_{ac}|$, the two eigenvalues coalesce to $e_{+}=e_{-}=0$ and, simultaneously, the corresponding eigenvectors merge, a hallmark of an exceptional point (EP). This spectral degeneracy is non-Hermitian in nature, distinct from Hermitian diabolic points where eigenvalues may coincide but eigenvectors remain orthogonal \cite{Heiss}.  

\begin{figure*}[htb!]
\centering
\includegraphics[width=0.4\textwidth]{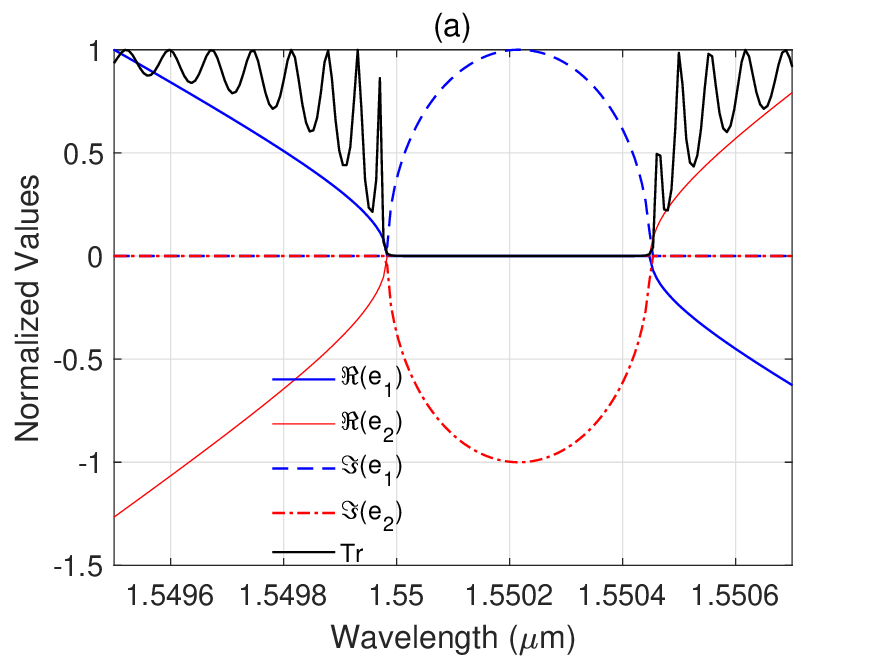}  
\includegraphics[width=0.4\textwidth]{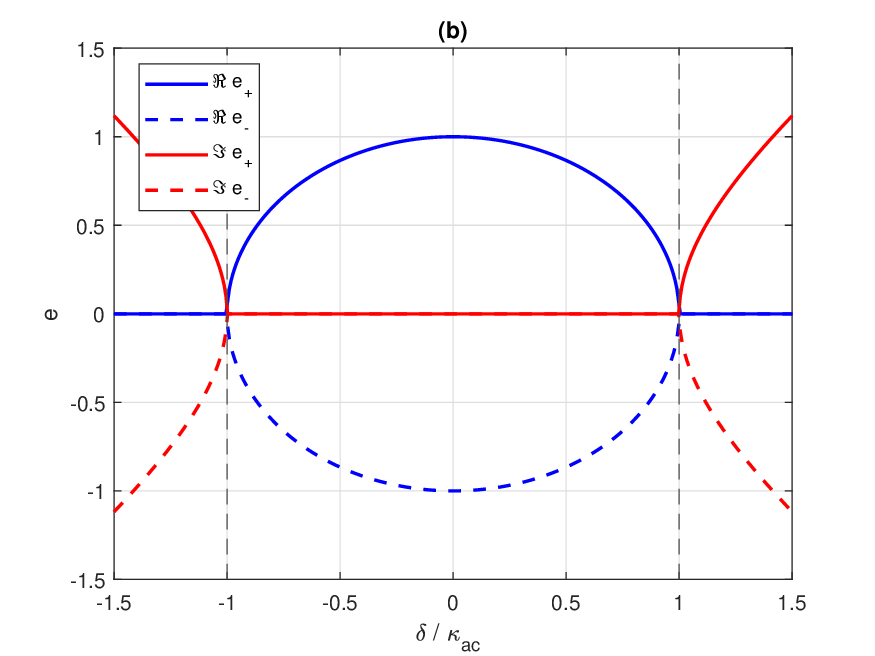} \\
\includegraphics[width=0.4\textwidth]{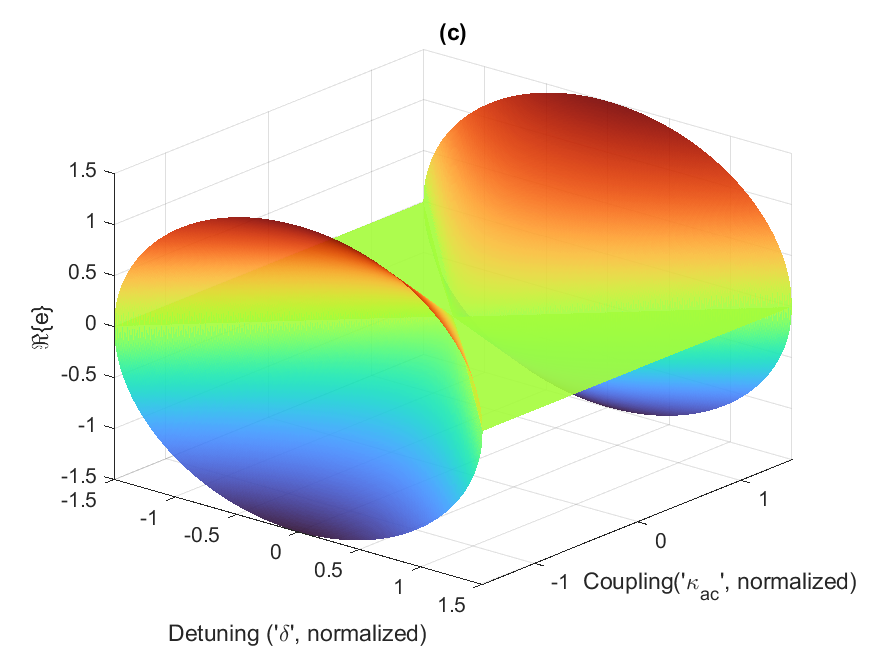}
\includegraphics[width=0.4\textwidth]{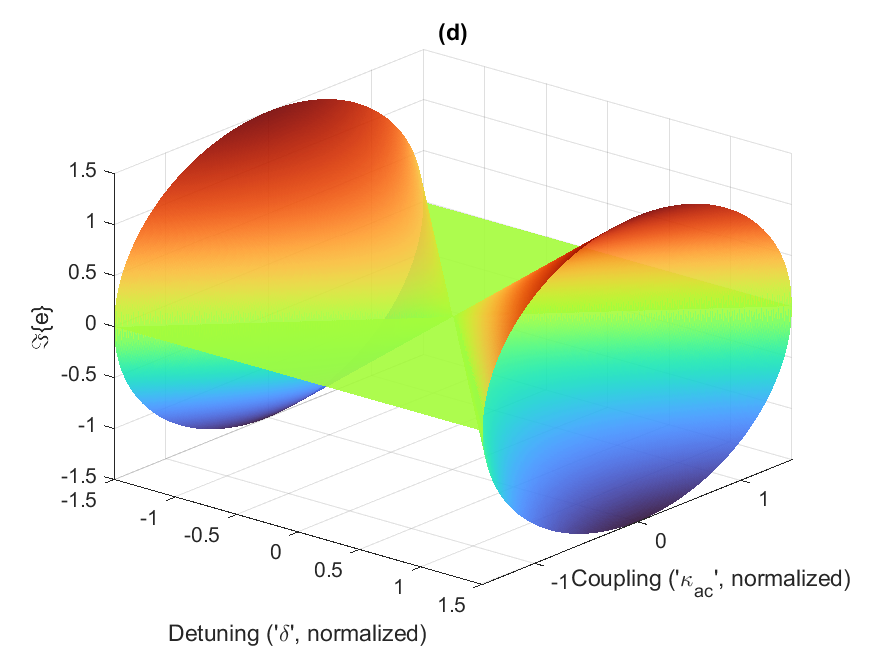} \\
\includegraphics[width=0.4\textwidth]{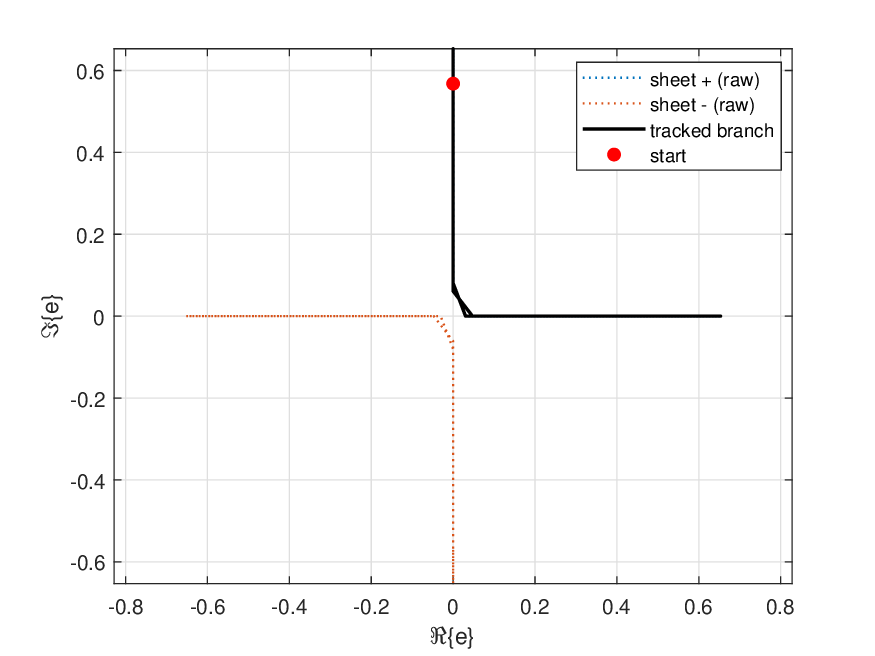}
\includegraphics[width=0.4\textwidth]{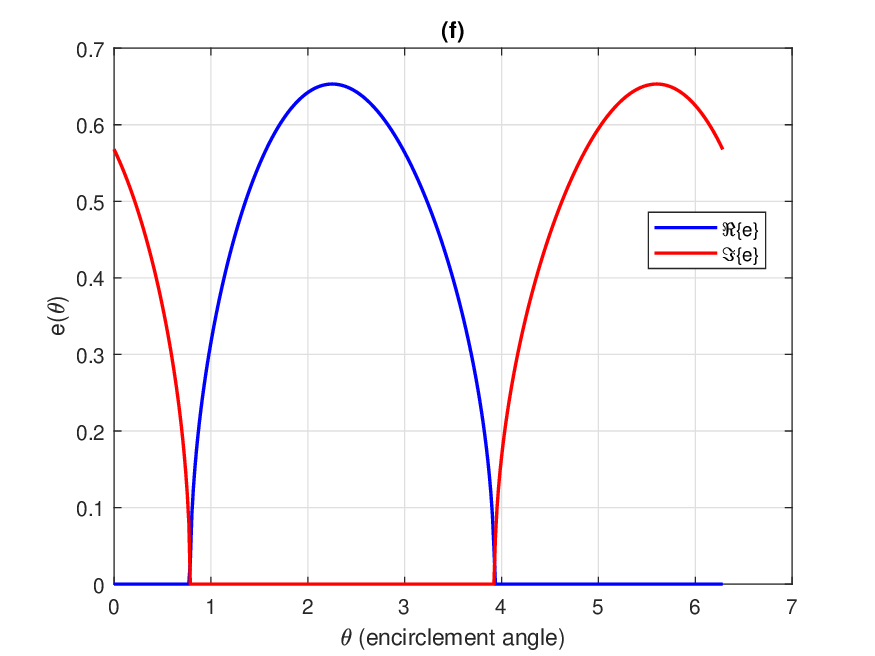}
\caption{\textbf{Topological nature of the exceptional point (EP) in the coupled-mode FBG system.} 
(a) Transmission spectrum of the FBG with eigenvalue coalescence at band edges (EPs). 
(b) One-dimensional eigenvalue transition showing real-to-complex bifurcation across the EP. 
(c–d) Riemann surfaces of $\Re(e)$ and $\Im(e)$ versus $(\delta,\kappa_{ac})$, showing the two-sheeted square-root topology and branch cut along $|\delta|=|\kappa_{ac}|$. 
(e) Encirclement of the EP in the complex plane, illustrating eigenvalue sheet exchange. 
(f) Evolution of $\Re(e)$ and $\Im(e)$ with encirclement angle $\theta$, confirming topological mode-switching behavior.}
\label{fig:EP_combined}
\end{figure*}

Figure~\ref{fig:EP_combined}(a) shows the transmission spectrum of the uniform FBG together with the calculated eigenvalues of $H$. The sharp turning points at the band edges are identified as EPs, where the eigenvalue trajectories meet, producing anomalous square-root dispersion. A one-dimensional parametric sweep across $\delta$ (Fig.~\ref{fig:EP_combined}(b)) makes this explicit: for $|\delta|<|\kappa_{ac}|$ both $e_{\pm}$ remain real, corresponding to propagating coupled modes, while at $|\delta|=|\kappa_{ac}|$ they coalesce into a defective eigenvalue, and for $|\delta|>|\kappa_{ac}|$ they bifurcate into complex-conjugate pairs, $e_{\pm}=\pm i \sqrt{\delta^{2}-\kappa_{ac}^{2}}$, indicating the onset of non-Hermitian splitting. This transition is the mathematical signature of a second-order branch point.  

To reveal the global topology, the eigenvalues were mapped as functions of both $\delta$ and $\kappa_{ac}$. The resulting surfaces for $\Re(e)$ and $\Im(e)$ (Fig.~\ref{fig:EP_combined}(c-d)) form a two-sheeted Riemann surface \cite{Farkas}. The branch cut along $|\delta|=|\kappa_{ac}|$ separates the real-valued and complex-valued domains, demonstrating the nontrivial square-root topology characteristic of EPs. Unlike a trivial avoided crossing, where two eigenvalue sheets merely repel each other, here the sheets are analytically connected through the branch cut, ensuring that analytic continuation across the cut permutes $e_{+}$ and $e_{-}$.  

This permutation is captured by parameter encirclement around the EP in the $(\delta,\kappa_{ac})$ plane. A circular loop parameterized as,
\begin{equation}
\delta(\theta) = \delta_{0} + r\cos\theta, \qquad 
\kappa_{ac}(\theta) = \kappa_{0} + r\sin\theta, \qquad \theta \in [0,2\pi],
\end{equation}
was used, where $(\delta_{0},\kappa_{0})$ denotes the EP location and $r$ is the radius of encirclement. Tracking the eigenvalues continuously along this path (Fig.~\ref{fig:EP_combined}(e-f)) shows that $e_{+}$ evolves smoothly into $e_{-}$ after one complete $2\pi$ rotation, and only after a second encirclement do the eigenvalues return to their original branches. This sheet-exchange property is the canonical topological invariant of an EP, equivalent to a geometric phase of $\pi$ acquired by the eigenvectors. The complex-plane trajectories and the corresponding $\Re(e)$ and $\Im(e)$ vs.\ encirclement angle both confirm this nontrivial topology.  

Taken together, Figs.~\ref{fig:EP_combined}(a-f) provide a unified view of the EP in the FBG system: eigenvalue coalescence in the transmission spectrum, one-dimensional square-root transition across detuning, the two-sheeted Riemann surface topology, and topological mode switching under encirclement. Importantly, this demonstration arises in a passive FBG without engineered gain–loss contrast, in contrast to most prior EP studies in active photonic systems. The square-root dispersion ensures that infinitesimal perturbations in $\delta$ or $\kappa_{ac}$ near the EP map to disproportionately large changes in $e_{\pm}$, underlining the physical basis for enhanced sensing. Moreover, the encirclement topology ensures robustness of this response against perturbation path variations, offering a new route to topologically-protected, ultra-sensitive fiber sensors.

\subsection{Merging FBG Band-edges with EPs}
Interestingly, \(|\kappa_{ac}| = \pm \delta\) also corresponds to the band-edge of the reflection spectrum in FBGs. In conventional Hermitian systems, band-edges define the transition between high- and low- transmission regimes. However, the band-edges are often not very sharp, and the transition from high- to low- reflectivity is often very smooth. 
To fully appreciate the effect of EPs on FBGs sensing characteristics, an extremely sharp-edged reflection spectrum for the FBG is required.  
In transfer matrix formalism, a sharp band edge occurs when the reflection coefficient transitions rapidly from nearly zero to nearly one.

For a steep band edge, the grating must reflect nearly all the incident light at the Bragg wavelength, leading to a high reflectivity and a sharp transition between high and low reflection. This is ensured when \( \kappa_{ac} L\gg 1\), which effectively means a strong coupling over a long grating. 
Increasing the grating length $L$ would enhance the reflectivity and make the band edges sharper. 

Another very important parameter dictating the sharpness of the bandedges is the group delay. The group velocity (\( v_g\)) represents how quickly information propagates through the grating. Mathematically, it is expressed as \( v_g = \frac{c}{n_g}\), where $n_g$ is the group index given by \( n_g = n_{eff} + \lambda \frac{d n_{eff}}{d\lambda}\). 
At the band edge, the dispersion relation \(\omega(\beta)\) becomes nearly flat, meaning \( \frac{d\beta}{d\omega} \rightarrow \infty \implies v_g \rightarrow 0\).  
This slow light effect enhances interaction time, leading to a sharper band edge. The stronger the coupling ($\kappa_{ac}$), the sharper this effect, leading to steeper band-edges.

\subsection{Grating Parameters}%Numerical Considerations}
To build upon the above discussion, in our analysis, we consider FBGs inscribed in telecommunication grade single-mode optical fiber, G.652 (11/16) \cite{G652}. The core (cladding) diameter is taken as 8.2 $\mu$m (125 $\mu$m) and the core (cladding) composition as GeO$_2$-doped SiO$_2$ (fuzzed SiO$_2$).

\textcolor{blue}{}

\begin{figure*}[htb!]
  \centering
  \includegraphics[width=8cm]{Tr_Re_e_Im_e_Temp.eps}
\includegraphics[width=8cm]{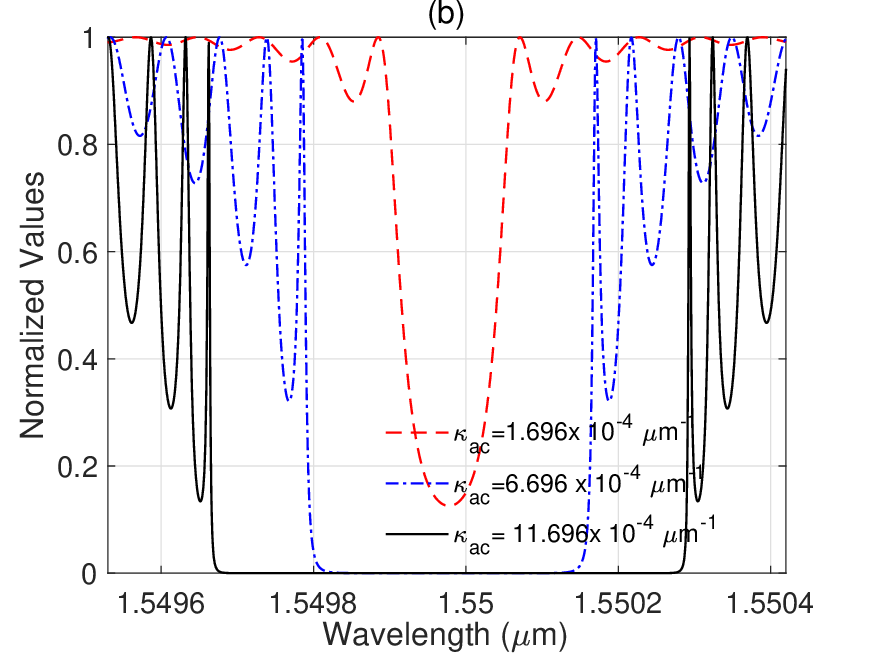}
\includegraphics[width=8cm]{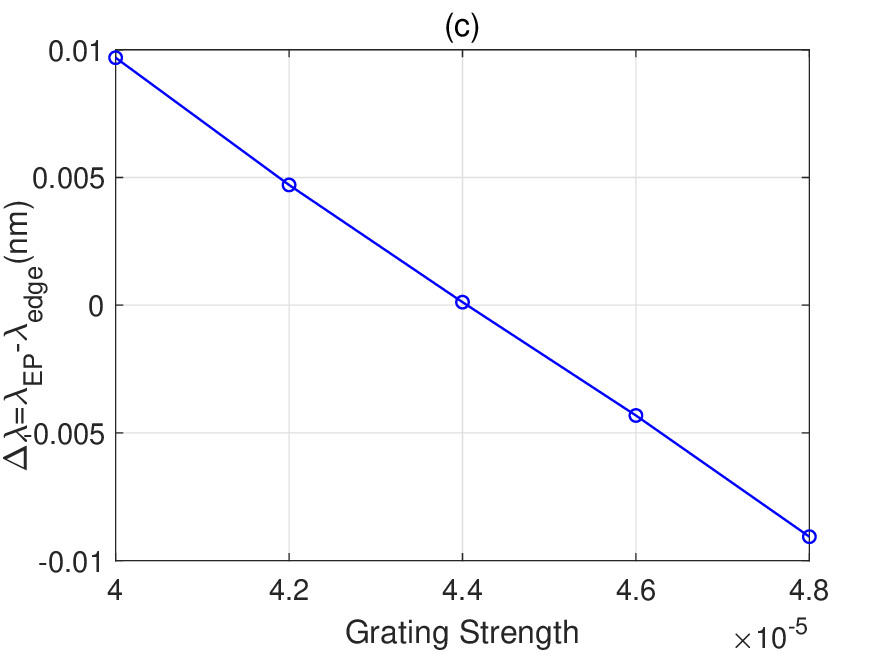}
\includegraphics[width=8cm]{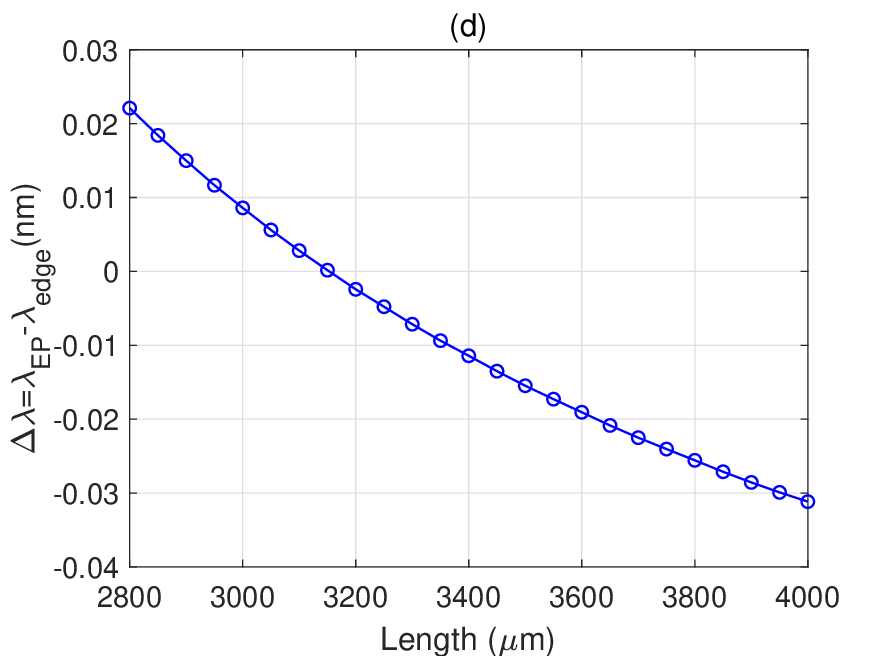}
\caption{(a) Transmission spectra (solid black curve) and spectral variation of $\Re{(e_1)}$ (solid blue curve), $\Re{(e_2)}$ (solid red-curve), $\Im{(e_1)}$ (dashed blue-curve) and $\Im{(e_2)}$ (dash-dotted red curve). 
(b) Transmission spectra for $\kappa_{ac} =1.696\times10^{-4}  \mu m^{-1}$ (dashed red curve), $\kappa_{ac} = 6.696\times10^{-4}  \mu m^{-1}$ (dash-dotted blue curve) and  $\kappa_{ac} = 11.696\times10^{-4} \mu m^{-1}$ (solid black curve). Wavelength shift of EP w.r.t. the (c) Grating strength (d) Length of the FBG. }
\label{EP}
\end{figure*}

\section{Results and discussion}
\subsection{\textit{Transmission spectra near EPs}}
The transmission spectra shown in Fig.~\ref{EP}(a) %derived from (9) 
is given by,
\begin{equation}
    t^2 = 1-\abs{r^2}
\end{equation}
\begin{equation}
    t^2 = 1-\frac{\abs{\kappa^2}\sinh^2{(\lambda L)}}{\abs{\kappa^2}\cosh^2{(\lambda L)}-\delta^2}
\end{equation}
In Fig.~\ref{EP}(a) the edges near EPs are defined as the 'band edges', which are described as the points located at the boundaries of the band gap. Near the band edges, the amplitudes A(z) and B(z) increase and decrease exponentially along the direction of the z-axis. Transmissivity at these band edges is 
\begin{equation}
    t_{b}^2 = 1-r_{b}^2
\end{equation}
which gives, 
\begin{equation}
    t_{b}^2 = 1- \frac{(\kappa L)^2}{1+(\kappa L)^2}
\end{equation}
the occurrence of these band edges appears at the wavelength,
\begin{equation}
    \lambda_b = \lambda_{max}\pm\frac{\overline{\delta n}}{2 n}\lambda_B
\end{equation}

%\section{Assessing final manuscript length}
%
\begin{figure*}[htb!]
  \centering
\includegraphics[width=8cm]
{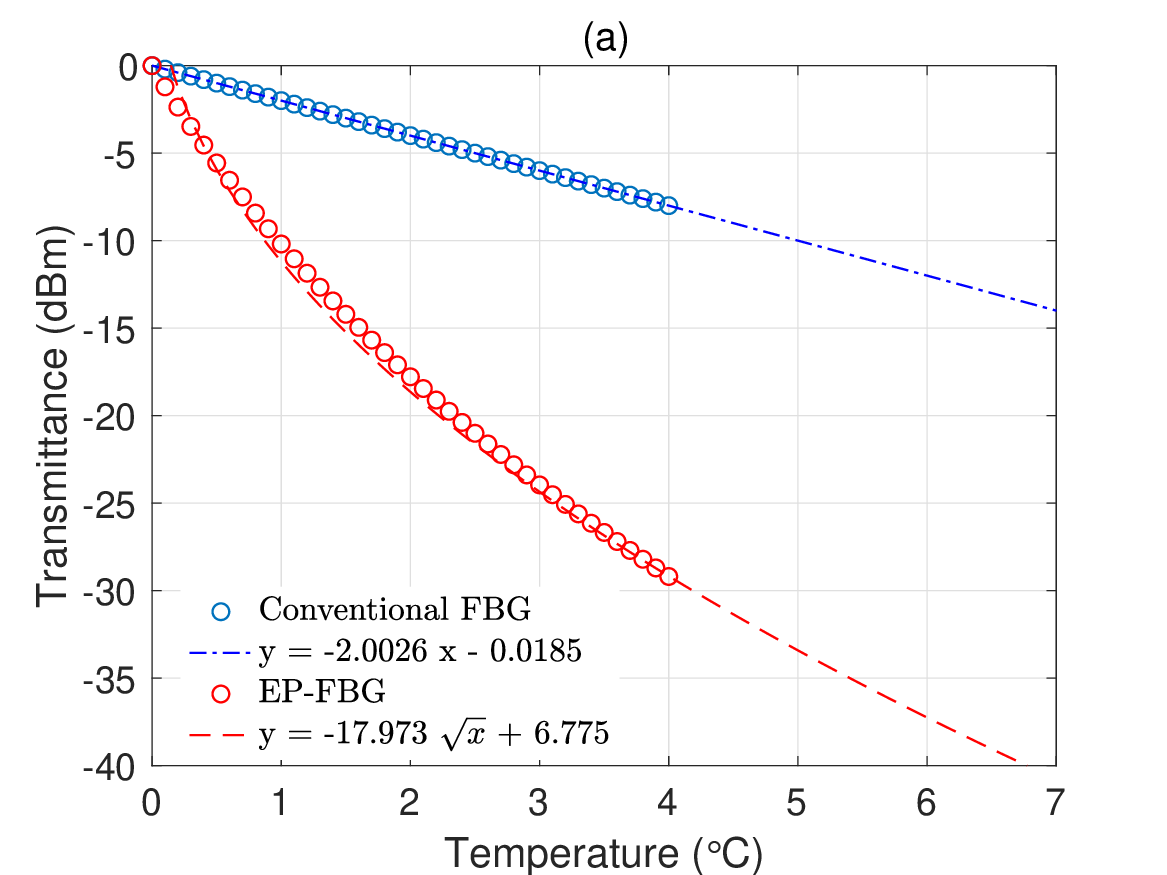}
\includegraphics[width=8cm]{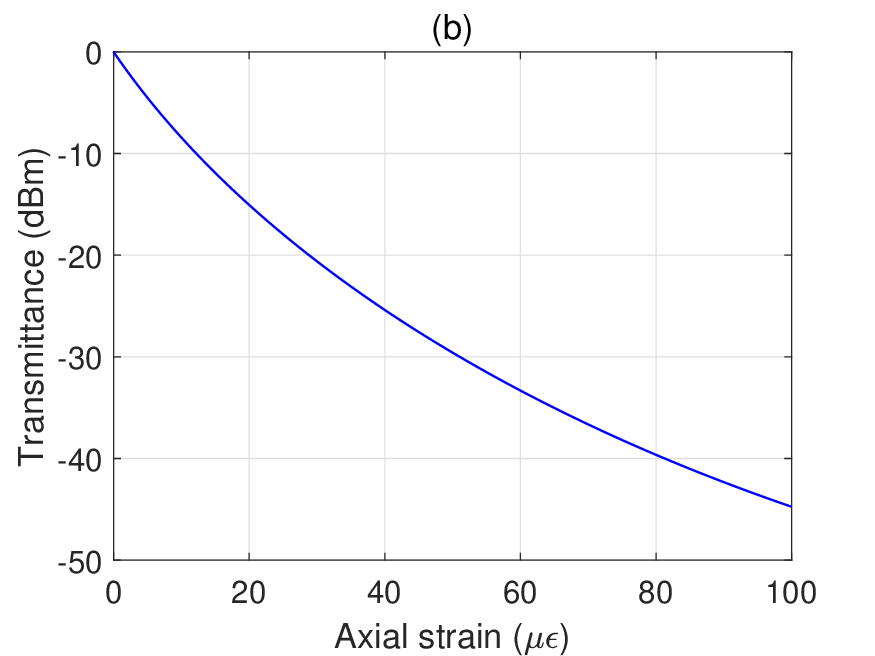}
\includegraphics[width=8cm]{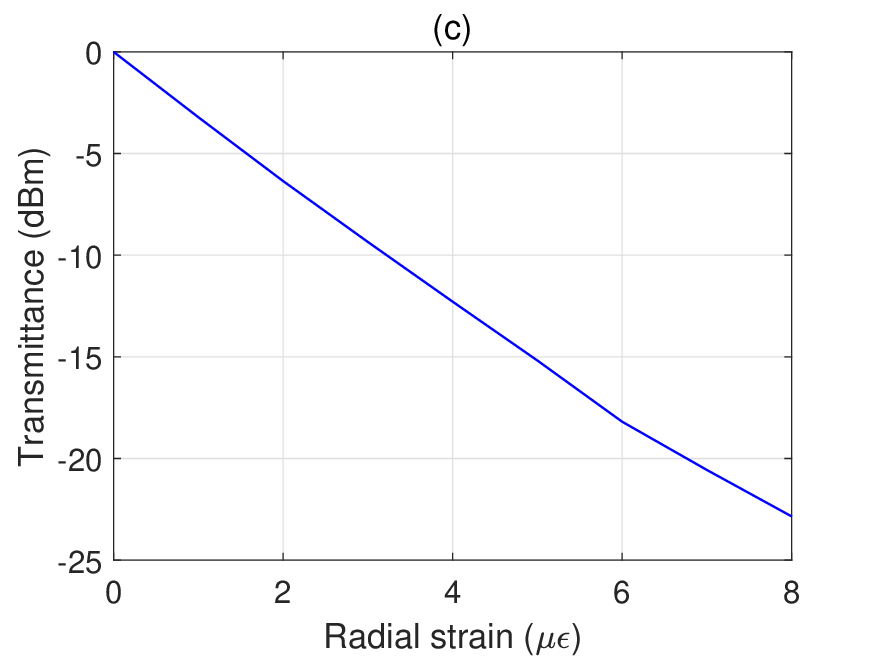}
\includegraphics[width=8cm]{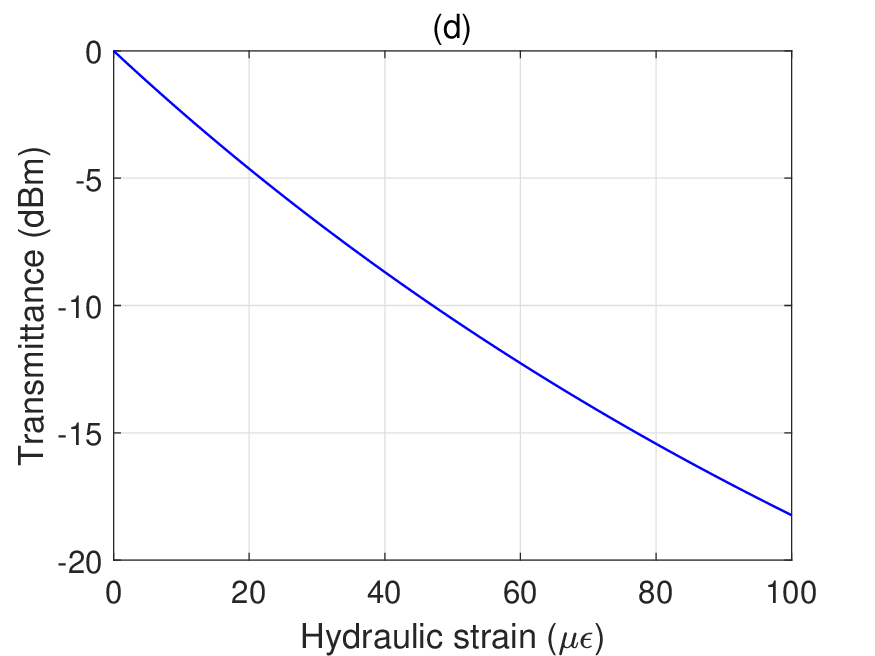}
\caption{Transmission spectra shift around the lower EP as a function of (a) temperature change at $\lambda$ = 1.54996 $\mu$m (dashed curve) and at $\lambda$ = 1.54994 $\mu$m (dash-dotted curve)
(b) axial strain, (c) radial strain, and (d) hydraulic strain. }
\label{EP_evolv}
\end{figure*}

In Fig.~\ref{EP}(a), we have plotted the transmission spectrum (solid black curve) of a uniform FBG centered at $\lambda_R = 1.55~\mu$m, showing a flattened reflection over a bandwidth of 0.36 nm. The grating length has been taken as 1 cm. The spectral variation of the real ($\Re{(e_{1,2})}$) and imaginary parts ($\Im{(e_{1,2})}$) of the eigenvalues of the degenerate guided modes are plotted using solid lines and dashed lines, respectively, in the same figure. The blue curves correspond to the forward propagating mode, and the red curves correspond to the backward propagating mode. We observe that at the reflection edges ($\sim$1549.9 nm and $\sim$1550.4 nm), the real and imaginary eigenvalues identically become zero. A closer observation of the eigenfunctions reveals that, along with the eigenvalues, the $|\psi_1 \rangle$ and $|\psi_2 \rangle$ coalesce at these unique wavelengths, rendering these two wavelengths associated with the exceptional points.\\
In Fig.~\ref{EP}(b), we have plotted transmission spectrum for three different coupling coefficients, namely $k_{ac}=1.696\times 10^{-4}~\mu m^{-1}$ (dashed curve), $k_{ac}=6.696\times 10^{-4}~\mu m^{-1}$ (dash-dotted curve) and $k_{ac}=11.696\times 10^{-4}~\mu m^{-1}$ (solid curve), respectively. As the coupling coefficient $\kappa_{ac}$ increases in a non-Hermitian system near an exceptional point (EP), the transmission spectrum becomes broader and steeper. This leads to higher sensitivity, as small changes in the system cause larger shifts in transmission. With further increase in $\kappa_{ac}$, the band edges of the transmission spectrum begin to coalesce with the exceptional points (EPs) of the system. 
In Fig.~\ref{EP}(c), we plot the difference between the spectral positions of the exceptional point and the band edge ($\Delta\lambda = \lambda_{EP} - \lambda_{edge}$), with increasing grating strength, $\sigma$. We observe that with increasing $\sigma$ the exceptional point moves closer to the band edge and eventually the $\Delta\lambda$ transitions from positive to negative value after $\sigma = 4.4 \times 10^{-5}$. The grating length has been fixed at 1 cm. At zero crossing, the EP merges with the band edge. 
Similarly, in Fig.~\ref{EP}(d), we plot the difference between the spectral positions of the exceptional point and the band edge. With increasing grating length, the exceptional point once again moves closer to the band edge and eventually $\Delta\lambda$ transitions from positive to negative for $L \sim 3160~\mu m$. 

\begin{figure*}[htb!]
  \centering
\includegraphics[width=8cm]{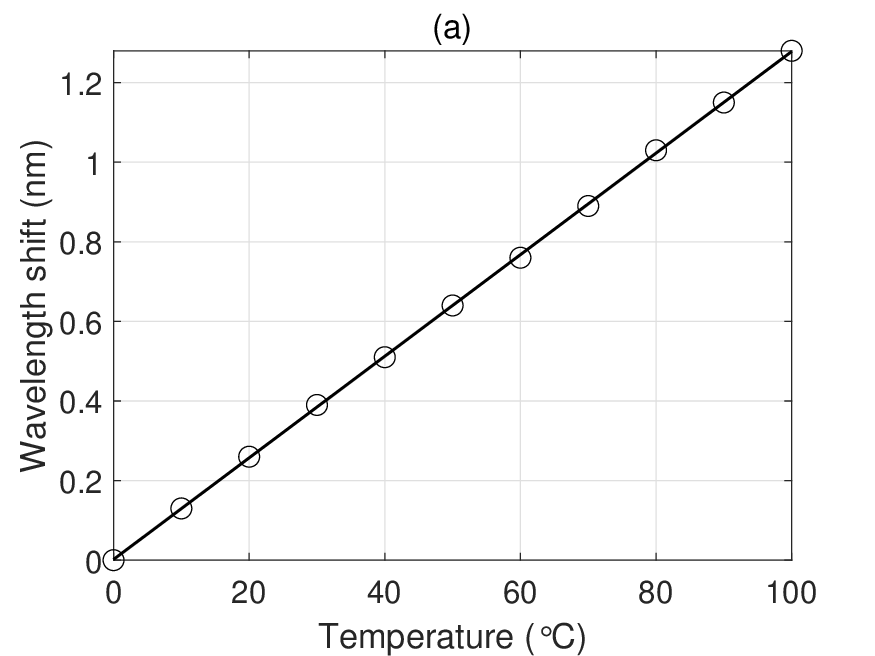}
\includegraphics[width=8cm]{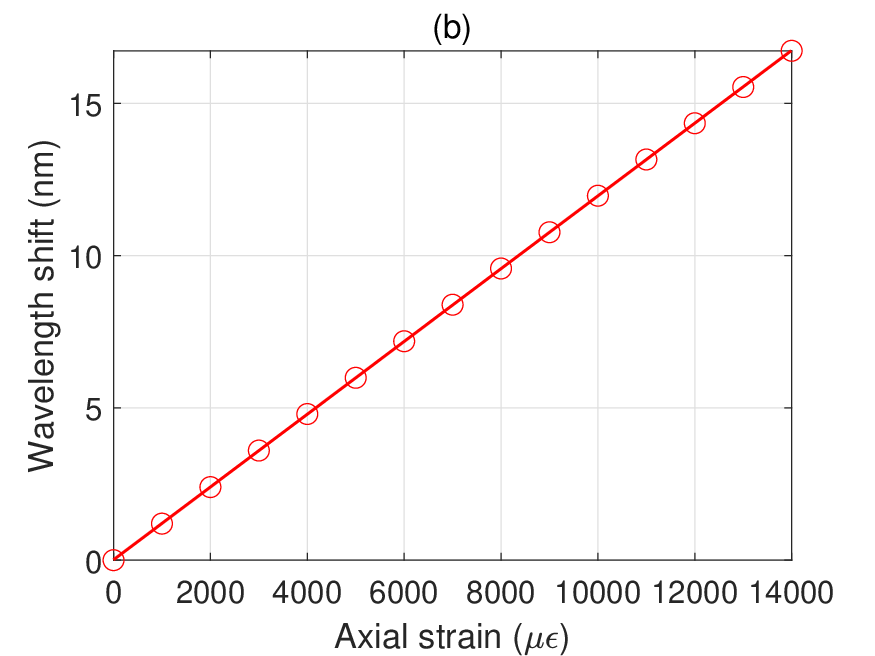}
\includegraphics[width=8cm]{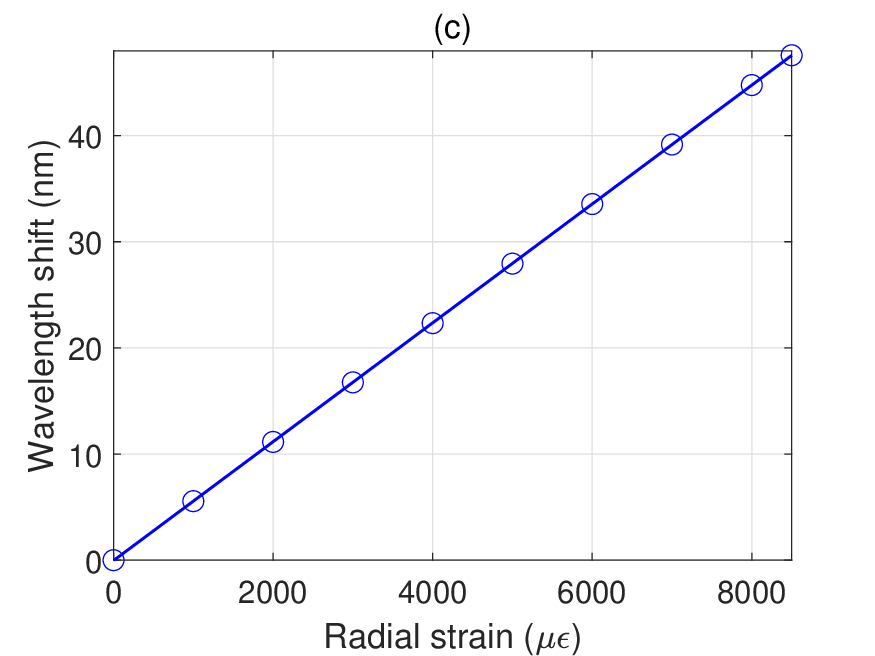}
\includegraphics[width=8cm]{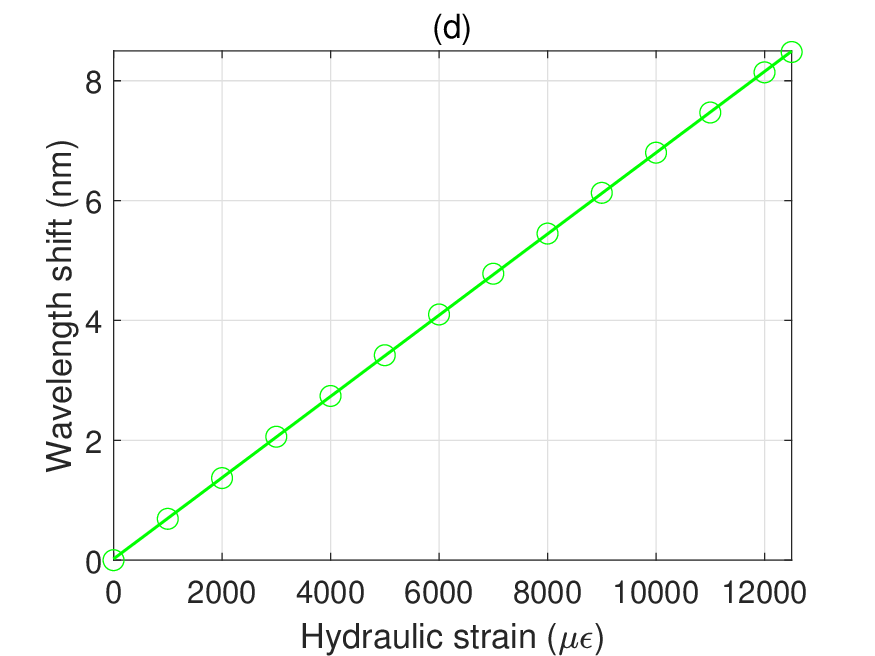}

\caption{Spectral shift around the lower EP as a function of (a) temperature (b) axial strain (c) radial strain and (d) hydraulic strain. }
\label{spect}
\end{figure*}
Operating the FBG close to its EPs is crucial as it dramatically elevates the sensor's sensitivity. 
In conventional FBGs, the reflection spectrum shifts linearly with changes in external parameters, e.g., strain or temperature variations, governed by well-established relationships involving the fiber’s effective refractive index and grating period. While this linearity is advantageous for simplicity and predictability, it imposes a strict limit on the sensitivity achievable with conventional FBG designs. Near EPs, however, the FBG’s response departs from this linear regime, exhibiting a pronounced nonlinear enhancement (cf. Fig.~\ref{EP_evolv}(a)) due to the coalescence of eigenmodes. This amplification arises because small perturbations in the FBG’s opto-geometric parameters induce disproportionately large shifts in the eigenvalues, translating to enhanced changes in the reflected light’s properties. By carefully tuning the FBG to operate near these critical points, the FBG becomes exquisitely sensitive to environmental fluctuations, drastically reducing the detection limit.

\subsection{\textit{Temperature Sensing}}
Temperature cross-sensitivity is a major source of inaccuracy in optical biosensors. Hence, evaluating the temperature sensitivity of the proposed sensor is crucial. Variations in temperature affect both: the core and cladding radii, as well as their refractive indices, which are determined using the following relation, \cite{smt_sms_jlt},
\begin{equation}
    a=a_0(1+\alpha\Delta T)
\end{equation}
\begin{equation}
    n_j = n_{0j}+\frac{dn_{0j}}{dt}\Delta T
\end{equation}
Here, $\Delta T=T_f-T_i$ is the change in temperature relative to the reference temperature, where $T_i$ and $T_f$ are initial and final temperatures, respectively.
In our analysis, both the core and cladding are assumed to exhibit identical thermal expansion coefficients of 
$5\times10^{-7}/^\circ$C, consistent with ITU-T G.652 standards. However, their thermo-optical coefficients differ, with values of $1.55\times10^{-5}/^\circ$C (core) and $1.06\times10^{-5}/^\circ$C (cladding) \cite{smt_sms_jlt}.
%In the present sensor design, the core and cladding exhibit identical thermal expansion coefficients, each valued at $5\times10^{-7}/^\circ$C, \textcolor{blue}{the thermo-optic coefficients of the core and cladding are taken as $1.55\times10^{-5}/^\circ$C and $1.06\times10^{-5}/^\circ$C, respectively \cite{smt_sms_jlt}.} Owing to their uniform thermal properties.
\\
To evaluate the temperature sensing performance of the proposed sensor, in Fig.~\ref{EP_evolv}(a) we have plotted the transmitted power (in dBm) as a function of ambient temperature for both a conventional FBG (blue circles) and the EP-enabled FBG (red circles). The results reveal a markedly stronger temperature dependence in the EP-enabled configuration, where the transmitted power exhibits a pronounced, nonlinear decline with increasing temperature. Notably, the fitted curves confirm the characteristic square-root response expected near the exceptional point, underscoring EP-enabled FBG's enhanced sensitivity compared to its conventional counterpart.

\begin{comment}
, rising from approximately 0 dBm to –15.7 dBm over the range of 0$^\circ$C to 1.7$^\circ$C. This corresponds to a high temperature sensitivity of approximately –9.227 dBm/$^\circ$C (= 119.3 $\mu$W/$^\circ$C), suggesting that this wavelength lies in close proximity to an exceptional point (EP), where small thermal perturbations result in disproportionately large optical responses.
In contrast, the response at 1549.97 nm shows a relatively linear and moderate variation in transmittance, changing from 0 dBm to –12.63 dBm over the same temperature range, corresponding to a sensitivity of approximately 7.429 dBm/$^\circ$C. 
The sharp nonlinear response observed at 1549.96 nm confirms the suitability of this wavelength for ultra-sensitive temperature sensing applications, highlighting the critical role of precise spectral alignment in the design of high-performance fiber-optic sensors.

\end{comment}

%Using the commercially available optical spectrum analyzer, with a noise floor value of $\sim$10 pW, the proposed sensor can detect temperature changes as small as \textcolor{red}{1.1 $\times$ 10$^{-8}$$~^\circ$C,} which is the smallest limit of detection reported to date. The sensor may find applications in various fields, including space exploration and other delicate measurements that require temperature-related corrections, e.g., gravitational wave detection, where the measurement of minute temperature variations is critical. 

\subsection{\textit{Axial-Strain Sensing}} 
When the fiber is subjected to a uniform and compressive pressure in the axial direction, the variations in the core radius is given by, \cite{smt_sms_jlt}
\begin{equation}
    a=a_{o}+\frac{\delta a}{\delta l}\Delta l
\end{equation}
\begin{equation}
    n_j = n_{0j} +\frac{\delta n_{0j}}{\delta l}\Delta l=n_{0j}-\frac{n_{0j}^3}{2l}[p_{12}-\sigma(p_{11}+p_{12})]\Delta l
\end{equation}   
where $(\delta a/\delta l)\Delta l=-(a\sigma/l)\Delta l=-a\sigma\varepsilon$ and $p_{ij}$ is strain-optic coefficient for fused silica, where $p_{11}$ and $p_{12}$ are given by 0.12 and 0.27 respectively \cite{palik}. The Poisson's ratio ($\sigma$) is given by 0.17. Here $n_j$ represents core and cladding refractive indices of an SMF, $\Delta l$ represents the elongation applied to the fiber and $n_{0j}$ denotes the refractive index when no strain is introduced.
%Figure 3(b) illustrates the wavelength shift of the FBG structure with respect to different values of axial strain (in $\mu\epsilon$), showing a linear response.
The term \(\epsilon = \Delta l/l\) represents the axial strain, and the value of $l$ is taken as 1 cm.
%As the axial stress increases, the spectral response steadily shifts toward higher wavelengths. 

\subsection{\textit{Radial-Strain Sensing}} 
The radial strain refers to pressure or deformation applied perpendicular to the fiber axis, affecting the transverse dimensions of the fiber. This induces stress across the fiber cross-section, modifying $n_{eff}$ via the stress-optic effect. Although less commonly explored compared to axial strain, the radial strain also affects the FBG's response particularly in embedded environments.
When the fiber having the Young's modulus of elasticity \textit{Y} is subjected to a uniform and compressive pressure $\delta P_0$ in the radial direction, without any axial strain at its ends, the variations in the core radius is given by \cite{Huang}
\begin{equation}
%a=a_{o}-\frac{(1-\sigma)}{Y}\delta P_0
a=a_{o}\Bigg[1-\frac{(1-\sigma)}{Y}\delta P_0 \Bigg]
\end{equation}
The variation in the refractive index of both the core and cladding as a result of radial strain is given by \cite{Huang}
\begin{equation}
   \delta n_j = \frac{n_{0j}^3}{2Y}[(p_{11}+p_{12})(1-\sigma)-2\sigma p_{12}]\delta P_0
\end{equation}
%Figure 3(c) illustrates the wavelength shift of the FBG structure with respect to different radial strain values (in $\mu\epsilon_r$), showing a linear response.
Here, $\epsilon_r=\Delta l/l =2\sigma\delta P_0/Y$ is the radial strain and the value of \textit{Y} is taken as $7\times10^{10}N/m^2$ \cite{Huang}. 
%As the radial strain increases, the spectral response steadily shifts towards higher wavelengths. 

\begin{figure*}[hbt!]
  \centering
\includegraphics[width=5.5cm]{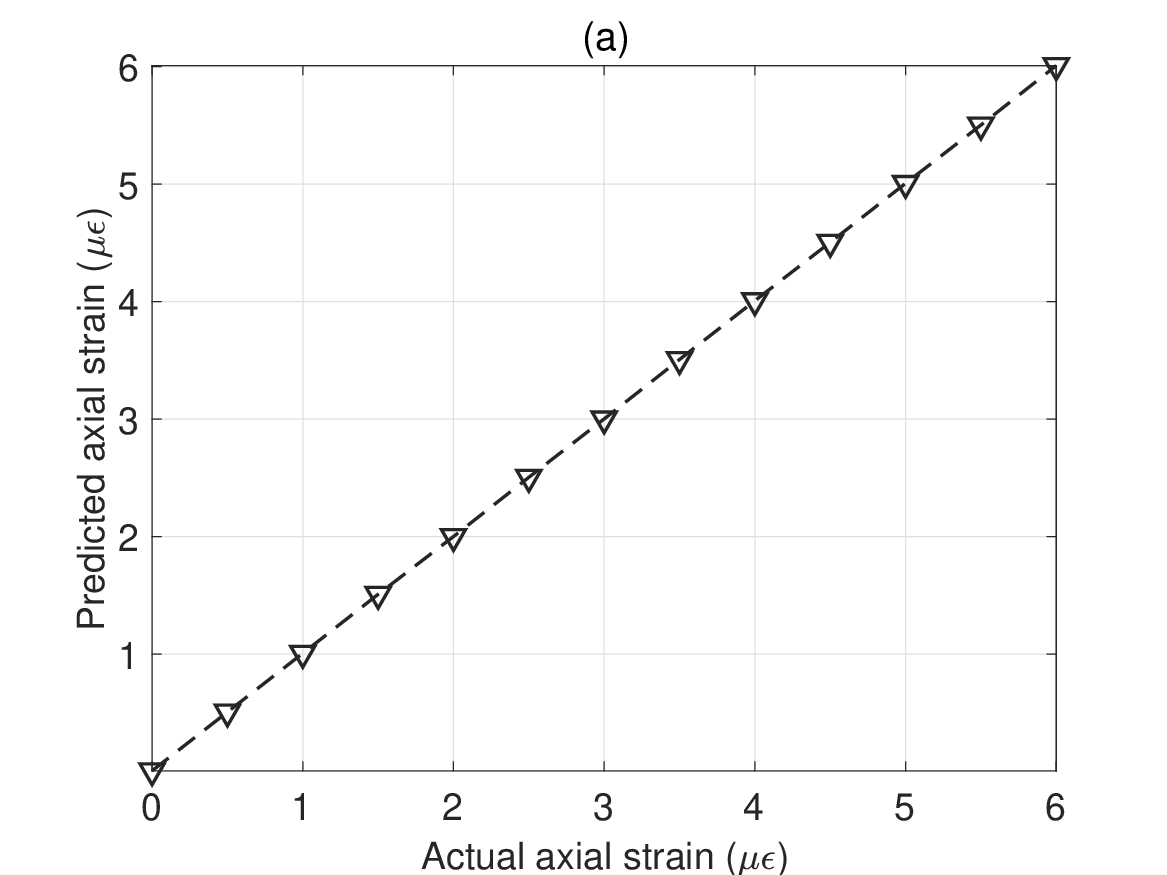}
\includegraphics[width=5.5cm]{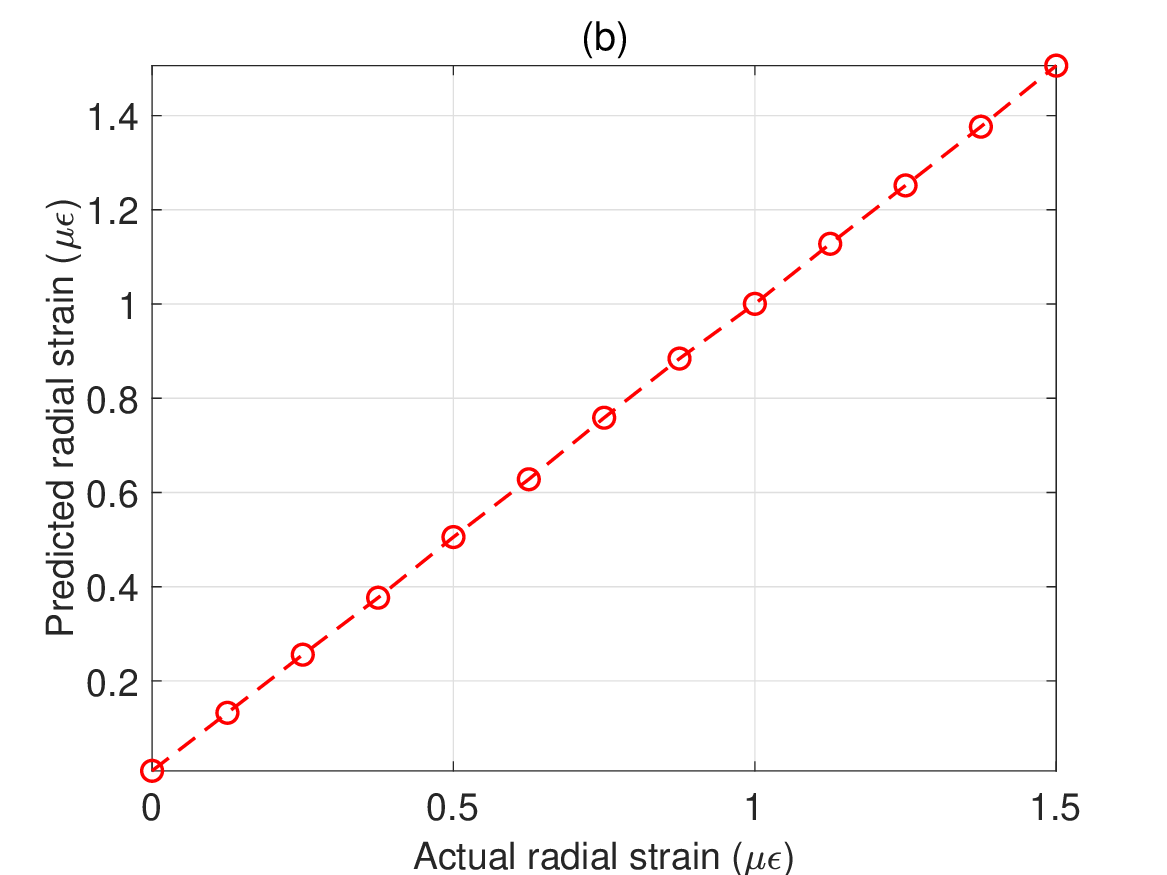}
\includegraphics[width=5.5cm]{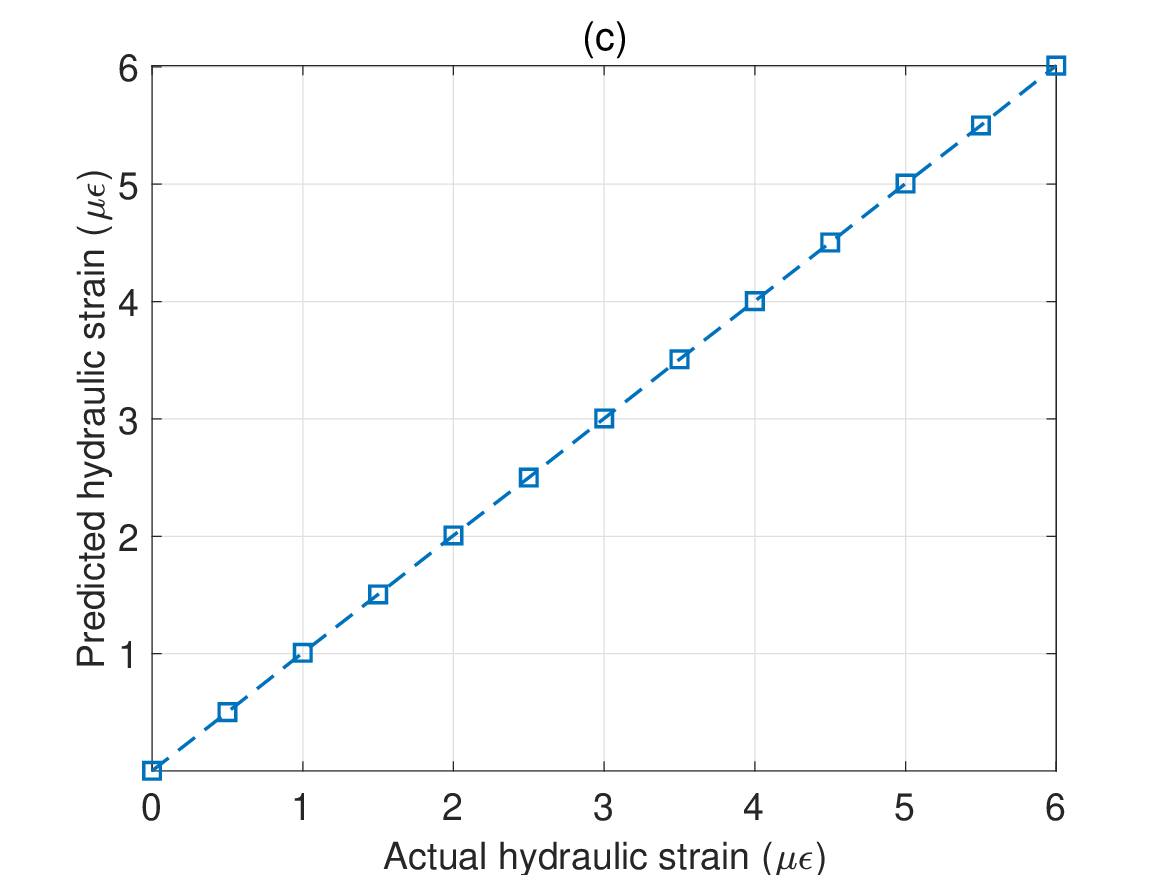}
\includegraphics[width=5.5cm]{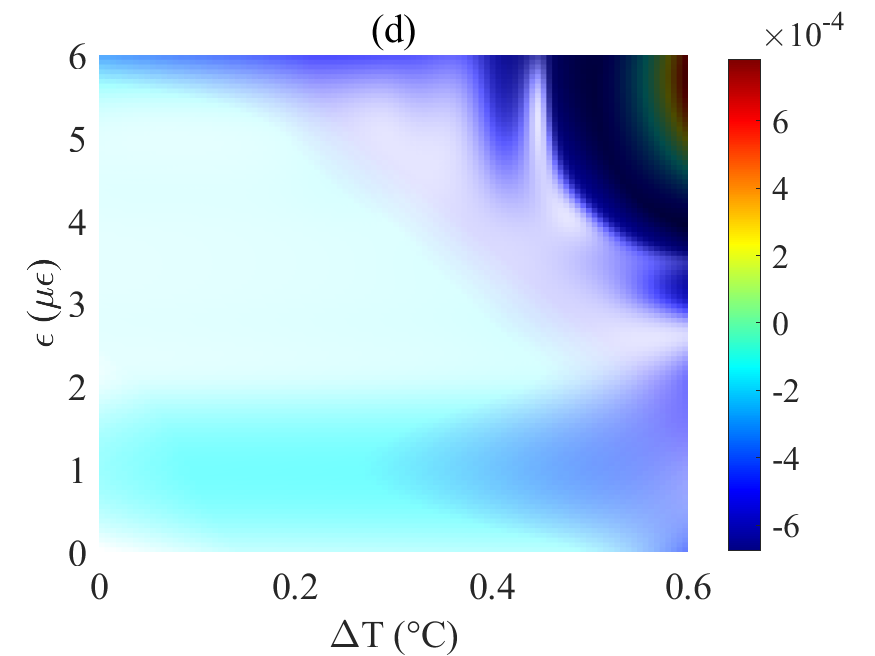}
\includegraphics[width=5.5cm]{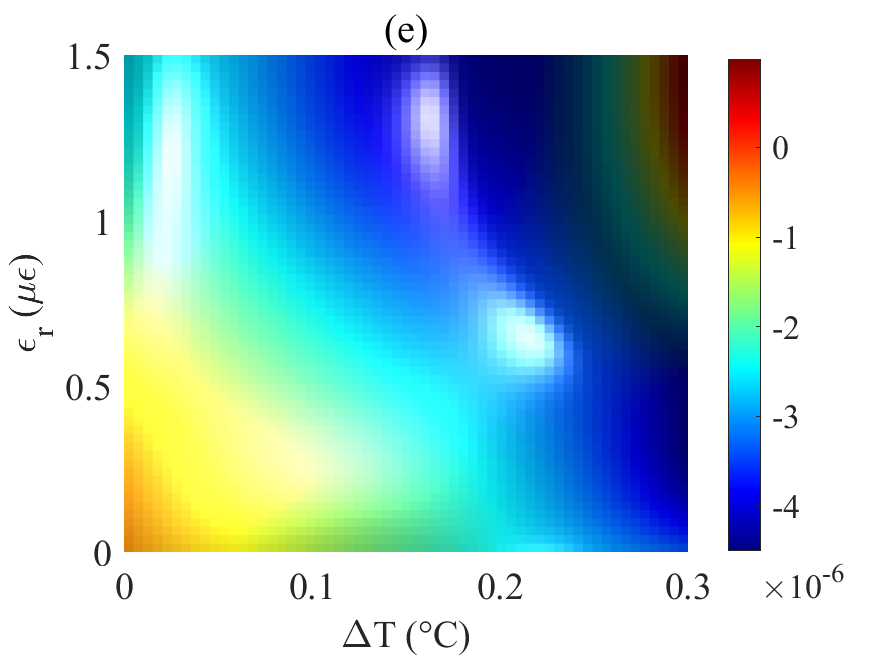}
\includegraphics[width=5.5cm]{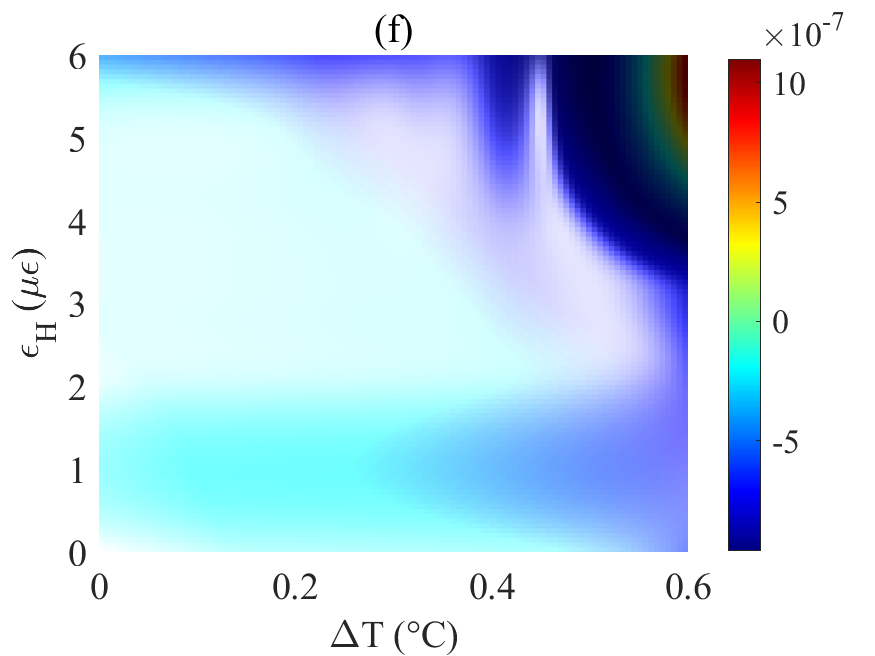}
\caption{Predicted vs actual parameters for (a) axial strain, (b) radial strain and (c) hydraulic strain. The temperature fluctuations have been considered in a range of 0-1 $^\circ$C. Similarity graph showing the error between the applied simultaneous perturbations and the estimated perturbations using spectral interrogation. Similarity graph showing the error between the applied simultaneous perturbations and the estimated perturbations using power interrogation. Simultaneous perturbations have been taken as temperature ($\Delta T$) and (d) axial-strain ($\Delta \epsilon$) (e) radial strain ($\Delta \epsilon_r$) and (f) hydraulic strain ($\Delta \epsilon_H$).}
\label{Error}
\end{figure*}

\subsection{\textit{Hydraulic-Strain Sensing}}
The hydraulic strain induced effect is observed when the FBG is subjected to isotropic external pressure, such as in a fluidic or underwater environment. Hydraulic strain sensing essentially relies on converting mechanical stress into pressure changes in a fluid. When the sensor is subjected to uniform stress due to the applied compressional pressure $\delta P_0$ and the axial pressure at the ends of the fiber is $-\delta P_0$, the resulting core diameter is expressed as, \cite{Othonos},
\begin{equation}
a=a_0\Bigg[1-(1-\sigma)\frac{\delta P_0}{Y} \Bigg].
\end{equation}
The corresponding change in refractive index of the fiber core and cladding regions is given by, \cite{Othonos}
\begin{equation}
   \delta n_j = \frac{n_{0j}^3}{2Y}(1-2\sigma)(p_{11}+2p_{12})\delta P_0 
\end{equation}
%Wavelength shifts corresponding to different hydrostatic strain values (in $\mu\epsilon_H$) are plotted in Fig. 3(d), showing a linear response.
Here, $\epsilon_H=\Delta l/l=-(1-\sigma)\frac{\delta P_0}{Y}$ is the hydraulic strain. 
%As the hydrostatic strain increases, the spectral response steadily shifts towards higher wavelengths. 

In Figs.~\ref{EP_evolv}(b)-(d) we have plotted the transmitted power shift around the lower EP as a function of axial strain, radial strain, and hydraulic strain, respectively. The respective sensitivities being 0.5 dBm/$\mu\epsilon$, 2.9 dBm/$\mu\epsilon$, and 0.2 dBm/$\mu\epsilon$. The distinct slopes observed in these figures arise from different perturbation–mode coupling mechanisms. Axial strain modifies both grating period and effective index via the photoelastic effect, producing the largest slope. In contrast, radial and hydraulic strains induce transverse stresses with weaker overlap with the guided mode, thereby reducing the effective coupling coefficient (see Eqs. (10)-(14)).

Using the commercially available optical spectrum analyzer, with a noise floor value of $\sim$10 pW, the proposed EP-enabled sensor can detect temperature changes as small as 1.1 $\times$ 10$^{-8}$$~^\circ$C, which is the smallest limit of detection reported to date. In Table-I, we have compared the sensitivities and estimated limit of detection (LOD) of the proposed EP-enabled FBG with the conventional FBG. A clear performance distinction emerges between the two FBGs, with the EP-enabled FBG outperforming the traditional FBG, offering a compelling route for next-generation, ultra-sensitive fiber-optic sensors. The sensor may find applications in various fields, including space exploration and other delicate measurements that require temperature-related corrections, e.g., gravitational wave detection, where the measurement of minute temperature variations is critical.

\begin{table*}[hbt!]
    \centering
    \caption{Performance Comparison of EP-enabled FBG and Conventional FBG}
    \begin{tabular}{|l|c|c|c|c|c|c|}
    \hline
\multirow{2}{*}{\textbf{S. No.}} & \multirow{2}{*}{\textbf{Parameter}} & \multicolumn{2}{|c|}{\textbf{Sensitivity}}& \multicolumn{2}{|c|}{\textbf{Estimated LOD}}\\
%\hline
 & & \textbf{EP-enabled FBG} & \textbf{Conventional FBG} & \textbf{EP-enabled FBG} & \textbf{Conventional FBG} \\
        \hline
       1  & Temperature & 0.9 $mW/^\circ$C & 0.036 $mW/^\circ$C & 1.1 $\times~10^{-8}~^\circ$C & 2.8 $\times~10^{-6}~^\circ$C\\
          \hline
    2  & Axial strain & 0.11 $mW/\mu\epsilon$ & 0.01 $mW/\mu\epsilon$ & $9.1\times10^{-8}$ $\mu\epsilon$ &$10^{-6}~\mu\epsilon$ \\
           \hline
    3  & Radial strain & 0.5 $mW/\mu\epsilon$& 0.08 $mW/\mu\epsilon$ & $2\times10^{-8}~\mu\epsilon$ & $1.25\times10^{-7}~\mu\epsilon$ \\
            \hline
    4 & Hydraulic strain & 0.045 $mW/\mu\epsilon$ & 0.008 $mW/\mu\epsilon$ & $2.22\times10^{-7}\mu\epsilon$ & $1.25\times10^{-6}~\mu\epsilon$\\
    \hline
    \end{tabular}
\end{table*}

\begin{comment}

\begin{table*}[hbt!]
    \centering
    \caption{Performance Comparison of EP-enabled FBG and Conventional FBG}
    \begin{tabular}{|c|c|c|c|c|}
        \hline% \toprule
\textbf{S.no.} & \textbf{Type of Sensor} & \textbf{Parameter} &\textbf{Sensitivity} & \textbf{LOD}  \\
\hline
1. & & Temperature & 0.9 $mW/^\circ$C & 1.1 $\times~10^{-8}~^\circ$C \\
 & EP-enabled FBG & Axial strain & 0.11 $mW/\mu\epsilon$ & $9.1\times10^{-8}$ $\mu\epsilon$\\
%\midrule
 %&  & 0.1 $mW/\mu\epsilon$ & $10^{-10}$ $\mu\epsilon$&  \\
%\hline
  &  & Radial strain &  0.5 $mW/\mu\epsilon$ & $2\times10^{-8}~\mu\epsilon$ \\
%\midrule
 & & Hydraulic strain & 0.045 $mW/\mu\epsilon$ & $2.22\times10^{-7}\mu\epsilon$\\
\hline%\midrule
2. & & Temperature & 0.036 $mW/^\circ$C  &  2.8 $\times~10^{-6}~^\circ$C\\
& Conventional FBG & Axial strain & 0.01 $mW/\mu\epsilon$  & $10^{-6}~\mu\epsilon$  \\
%\midrule
 % &  &  0.1$mW/\mu\epsilon$ & & \\
  %\midrule
  &  & Radial strain & 0.08 $mW/\mu\epsilon$   & $1.25\times10^{-7}~\mu\epsilon$ \\
%\midrule
 & & Hydraulic strain & 0.008 $mW/\mu\epsilon$ & $1.25\times10^{-6}~\mu\epsilon$ \\
\hline
    \end{tabular}
    \label{tab:placeholder}
\end{table*}

\end{comment}

\subsection{Spectral Shift Behavior}

Figures~\ref{spect}(a)-(d) illustrate the spectral shift behavior of the FBG's spectrum in response to temperature, axial strain, radial strain, and hydraulic strain variations. We observe that across all four cases, the spectrum shows a linear red shift. 
The temperature induced shift (Fig.~\ref{spect}(a)) arises
primarily due to the thermal expansion of the fiber's cross sectional dimensions and the grating period and grating length, as well as the thermo-optic effect. The calculated slope corresponding to the temperature change is $78.43,\text{nm}/^\circ\text{C}$, over a temparature variaiton of $\Delta T = 100~^\circ\text{C}$.
%
%The sensitivity of the sensor is studied by observing how the spectral position of the exceptional points (EPs) shifted with changes in temperature and applied strain. In Fig. 3(a), a clear redshift in the EP wavelength is observed with increasing temperature, primarily due to the thermal expansion of the fiber and the thermo-optic effect. The calculated slope corresponding to the temperature change is $78.43,\text{nm}/^\circ\text{C}$.} %Similarly, Figs. 3(b), 3(c) and 3(d) show the wavelength shifts under axial, radial, and hydraulic strains, respectively. In each case, the response is linear, indicating stable and repeatable sensor behavior. The calculated slopes for axial, radial, and hydraulic strains are $836.17, \text{nm}/\mu\epsilon$, $178.684, \text{nm}/\mu\epsilon$, and $1471.752, \text{nm}/\mu\epsilon$, respectively.
Figure~\ref{spect}(b) illustrates the wavelength shift of the FBG with respect to different values of axial strain (in $\mu\epsilon$), showing a linear response. As the axial stress $(\epsilon)$ increases, the spectral response gradually shifts to higher wavelengths.
This axial strain induced spectral shift is primarily governed by the elongation of the grating period and the strain-induced modification of the refractive index through the photo-elastic effect (cf. eq.(12)).  
The calculated slope value for axial strain is $836.17,nm/\mu\epsilon$ over a range of 14000 $\mu\epsilon$.
Similarly, Fig.~\ref{spect}(c) illustrates the wavelength shift of the FBG with respect to different radial strain values (in $\mu\epsilon$). As the radial strain $(\epsilon_r)$ increases, the spectral response steadily shifts towards higher wavelengths. The calculated sensitivity for radial strain is $178.684,nm/\mu\varepsilon$, over a range of 8000 $\mu\epsilon$..
Finally, the wavelength shifts corresponding to different hydrostatic strain values (in $\mu\epsilon$) are plotted in Fig.~\ref{spect}(d). As the hydrostatic strain $(\epsilon_H)$ increases, the spectral response gradually shifts to higher wavelengths. The calculated sensitivity for hydraluic strain is $141.752,nm/\mu\epsilon$, over a range of 12000 $\mu\epsilon$.

We would like to mention here that a substantially larger range of external perturbations is accessible under spectral interrogation compared to power interrogation. This distinction arises due to the fundamental differences in the underlying sensing principles. Spectral interrogation techniques track the shift in the Bragg wavelength ($\lambda_B$) as a function of the external stimulus. In contrast, power interrogation involves monitoring the reflected or transmitted optical power at a fixed wavelength. As the external perturbation increases, the operating wavelength drifts across the Bragg reflection spectrum, causing the measured power (at a fixed wavelength) to initially increase toward a maximum, then decrease again toward a side lobe. Thus, beyond a certain stimulus threshold, further increase in perturbations results in power values corresponding not to the main lobe but to higher-order side lobes, leading to ambiguity in interpreting the sensor response. This intrinsic limitation restricts the usable dynamic range of power interrogation technique.

%These results confirm that the sensor is highly sensitive around the EPs, where even small changes in external conditions lead to significant spectral shifts, making it suitable for precise temperature and strain sensing.
\section{Simultaneous Measurement of Perturbations}
The FBG is highly sensitive to temperature fluctuations, which can arise from daily and seasonal changes, differences between indoor and outdoor environments, or localized heating and cooling. These variations contribute to the cross-sensitivity between strain and temperature. Temperature cross-sensitivity is one of the primary factors that can lead to false alarms. To overcome this challenge, certain designs have been developed to naturally eliminate phase shifts caused by temperature variations \cite{smt_sms_OL, smt_EColi_OL, smt_temp_SnB}, and/or proper device packaging with a suitable ceramic to compensate for thermal expansion of the optical waveguide \cite{patent}. Owing to the presence of two EPs of the current sensor, it is possible to carry out a simultaneous measurement of two perturbation parameters. 
The characteristics equation for the FBG is
\begin{equation}
    \delta = \kappa_{dc} + \beta-\frac{\pi}{\Lambda}.
\end{equation}
Here $\kappa_{dc}$ is the self-coupling coefficient and we have assumed the coupling between the forward and backward propagating core modes. Its value is typically very small, but it plays a crucial role in determining the resonance wavelength. The resonances occur when $\delta = 0$, and the EPs exist at $\delta = \kappa_{ac}$. The spectral shift of EP is influenced not only by applied strain but also by additional physical factors that affect both the refractive index and other parameters of the waveguide, e.g., changes in surrounding temperature. Using %Eq.~\ref{phase_matching}, 
Eq. 15 the shifts in resonance wavelength ($\lambda_R$) arising due to a change in the temperature ($\Delta T$) and the other perturbation parameter ($\Delta \chi$) is expressed as:
\begin{multline}
    \Delta \lambda_R = 2 \Bigg[\bigg(\frac{\partial n_{eff}}{\partial T }\Delta T + \frac{\partial n_{eff}}{\partial \chi }\Delta \chi\bigg)\Lambda+\\ n_{eff}\bigg( \frac{\partial\Lambda }{\partial T}\Delta T + \frac{\partial \Lambda}{\partial \chi}\Delta \chi\bigg)  \Bigg]
\end{multline}

Now, since $m$ independent, linear equations could be solved to obtain $m$ unknown parameters, tracing the spectral shift of two EPs the two independently varying perturbation parameters could be estimated as follows. The wavelength shifts of the two EPs, located at $\lambda_{EP1}$ and $\lambda_{EP2}$, due to changing temperature, $\Delta T$, and strain, $\Delta \chi$, are expressed as,
\begin{equation}
    \begin{bmatrix}
\Delta \lambda_{R1}\\
\Delta \lambda_{R2}
\end{bmatrix} =
\mathcal{M}
\begin{bmatrix}
\Delta T\\
\Delta \chi
\end{bmatrix} 
\label{simultaneous}
\implies
    \begin{bmatrix}
\Delta T\\
\Delta \chi
\end{bmatrix} =
\mathcal{M}^{-1}
\begin{bmatrix}
\Delta \lambda_{R1}\\
\Delta \lambda_{R2}
\end{bmatrix}, 
\end{equation}
where, $\mathcal{M}$ is the sensitivity matrix expressed as,
\begin{equation}
    \mathcal{M} = \begin{bmatrix}
\delta \lambda_{R1}/\delta T & \delta \lambda_{R1}/\delta \chi \\
\delta \lambda_{R2}/\delta T &\delta \lambda_{R2}/\delta \chi
\end{bmatrix}.
\end{equation}
\begin{figure*}[hbt!]
  \centering
  \includegraphics[width=5.5cm]{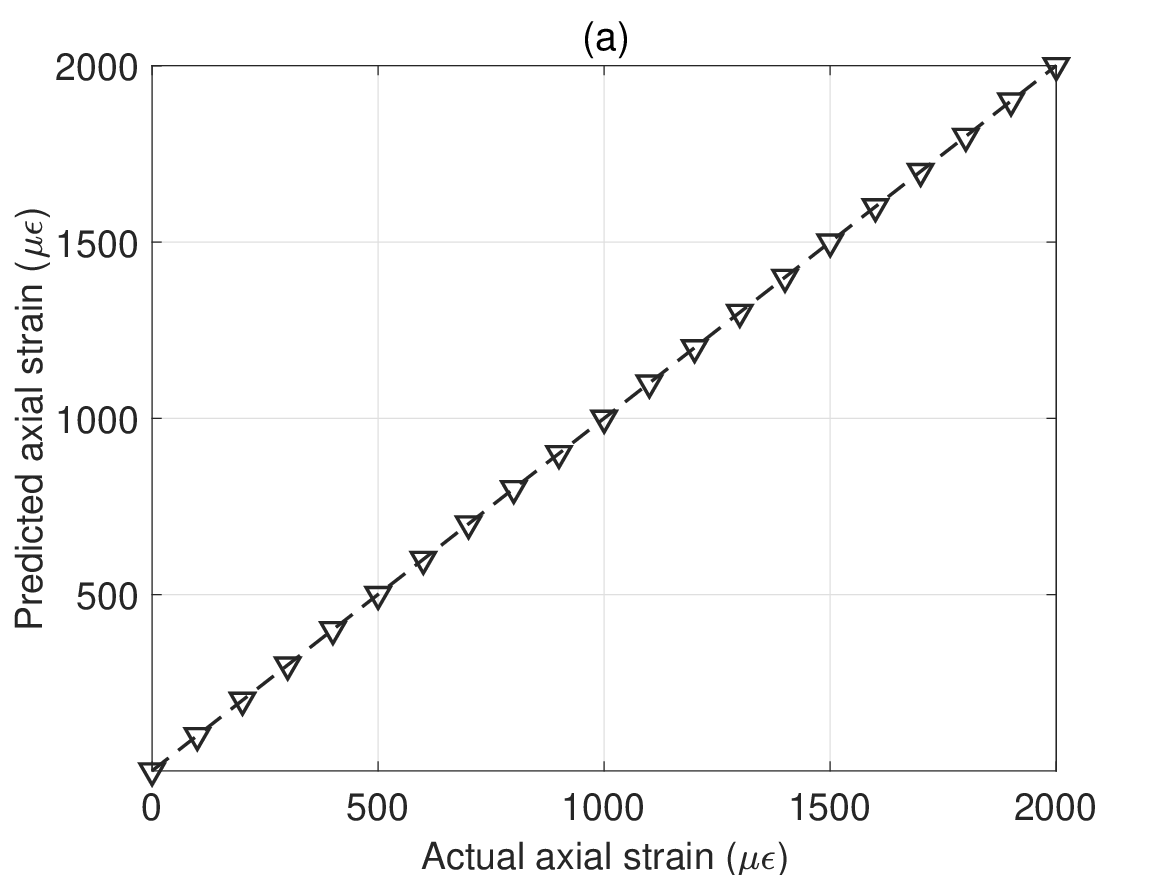}
\includegraphics[width=5.5cm]{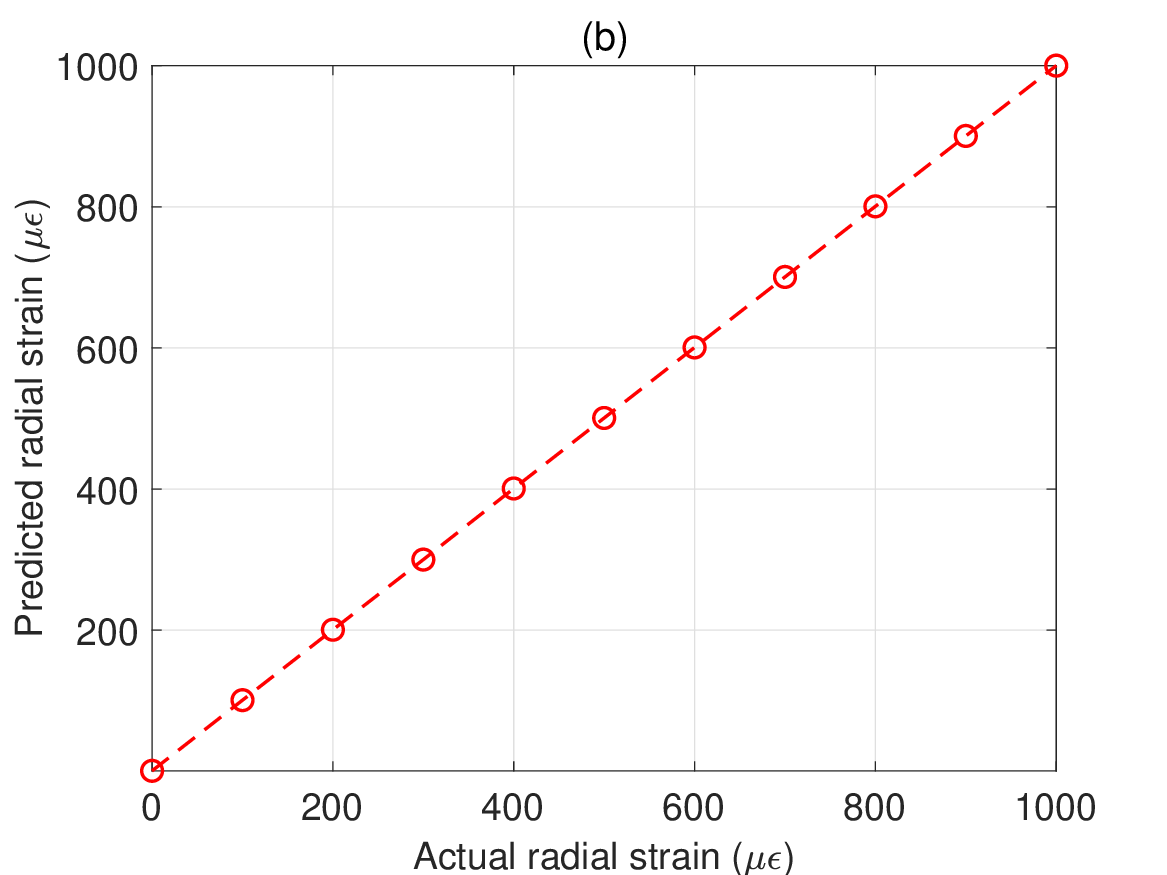}
\includegraphics[width=5.5cm]{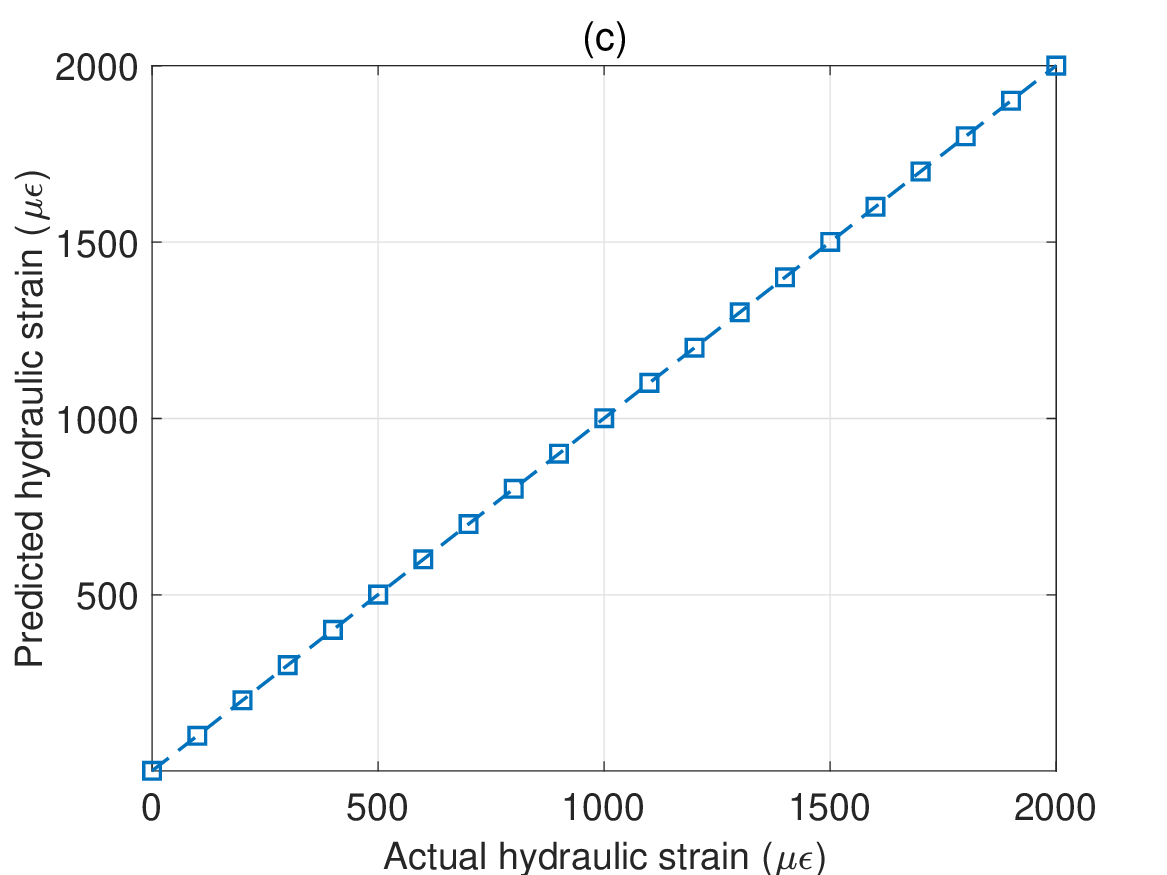}
\includegraphics[width=5.5cm]{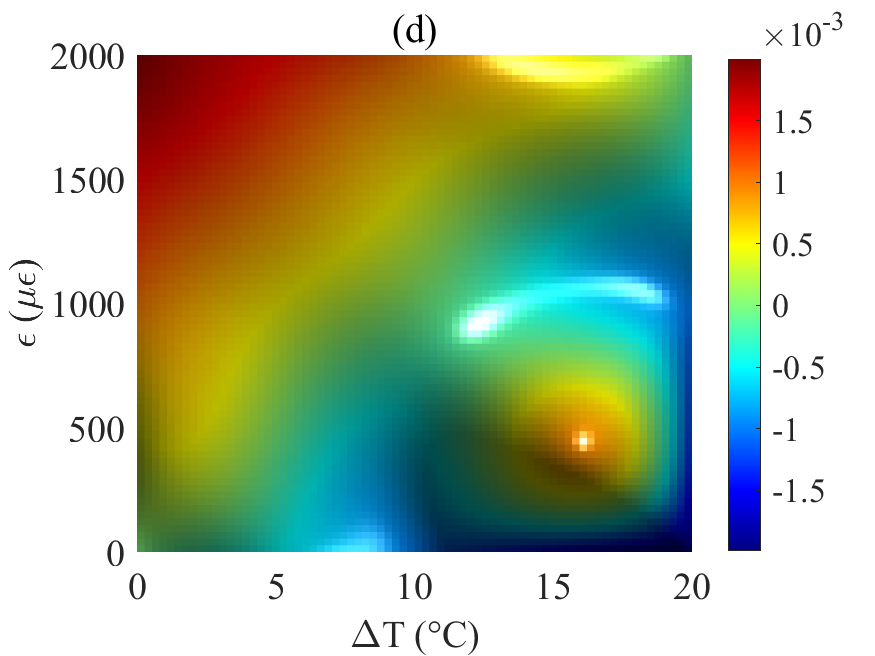}
\includegraphics[width=5.5cm]{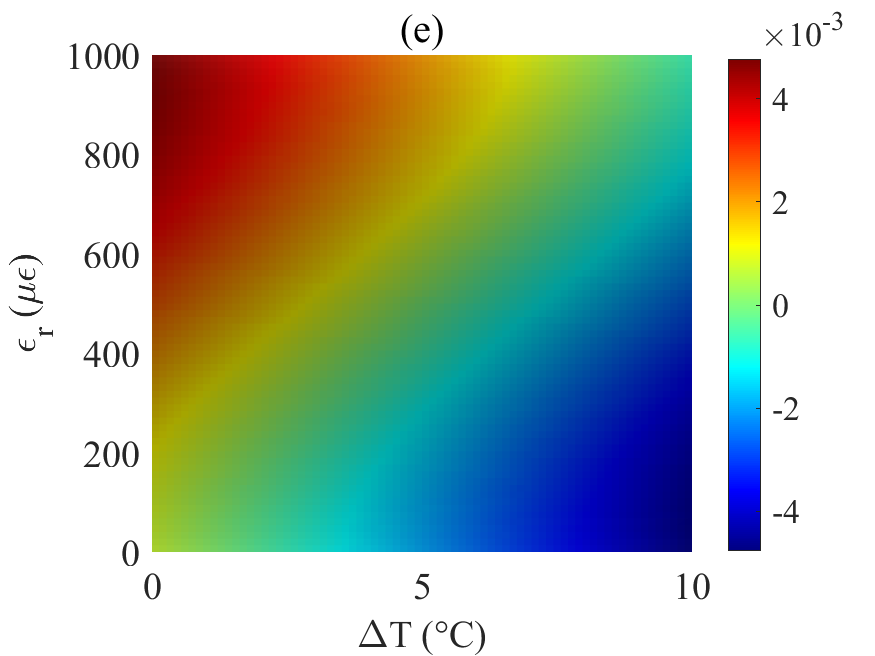}
\includegraphics[width=5.5cm]{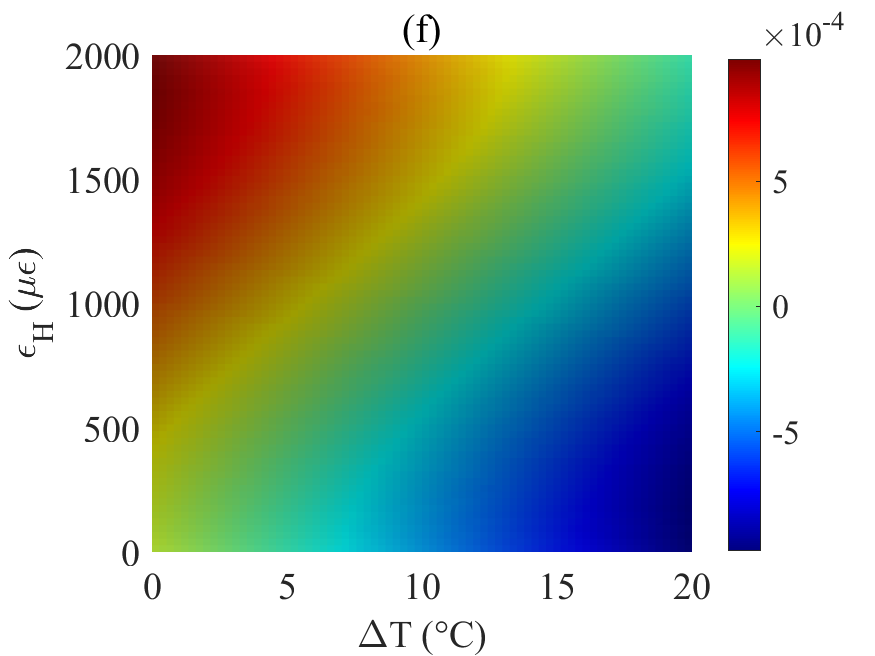}
\caption{Predicted vs actual parameters for (a) axial strain, (b) radial strain and (c) hydraulic strain. The temperature fluctuations have been considered in a range of 0-20 $^\circ$C. Similarity graph showing the error between the applied simultaneous perturbations and the estimated perturbations using spectral interrogation. Simultaneous perturbations have been taken as temperature ($\Delta T$) and (d) axial-strain ($\Delta \epsilon$) (e) radial strain ($\Delta \epsilon_r$) and (f) hydraulic strain ($\Delta \epsilon_H$).}
\label{Error1}
\end{figure*}

Thus, the two simultaneously changing perturbation parameters, $\Delta T$ and $\Delta \chi$, can be calculated by measuring the spectral shifts of the two EPs. Similarly, by appropriately defining the matrix $\mathcal{M}$, in terms of power variations to varying perturbations, and incorporating the concerning shifts on the $rhs$ of Eq.(18), the same method could be used for simultaneous measurements of two parameters using the power interrogation technique as well.

%The sensitivity matrix $\mathcal{M}$ (Eq. 18) was evaluated for condition number across the perturbation range, with its values largely remaining below 10 indicating numerical stability. For inversion, we have applied Tikhonov regularization to suppress amplification of measurement noise. Additionally, we also reconstructed strain variations in the presence of $\pm0.1~^\circ$C temperature fluctuations, observing accurate recovery of applied patterns with mean absolute error and root mean square error being $\sim0.3~\mu$m and $\sim0.35~\mu$m, respectively, over a strain variation of $\pm20~\mu$m.% (see supplementary figure S1).}}

\subsection{Tikhonov Regularization and Stability Analysis}

The recovery of physical parameters such as temperature and strain from the accessible observables at two exceptional points requires the inversion of a sensitivity matrix $\mathcal{M} \in \mathbb{R}^{2\times 2}$. In practice, this inversion is ill-conditioned: small perturbations in the measurement vector $y$ (due to noise or finite spectral resolution) can be strongly amplified, leading to unstable or non-physical parameter estimates. To suppress this instability, we employ Tikhonov regularization, which replaces the naïve least-squares inversion with a penalized optimization problem of the form, \cite{tikhonov},
\begin{equation}
x_{\lambda} = \arg\min_{x} \, \left\{ \| \mathcal{M}x - y \|_{2}^{2} + \lambda^{2} \| Lx \|_{2}^{2} \right\},
\end{equation}
where $x = [\Delta T, \, \Delta \epsilon]^{T}$ denotes the unknown physical perturbations, $L$ is the regularization operator (here taken as the identity), and $\lambda > 0$ is the regularization parameter controlling the balance between fidelity to the measured data and smoothness/ robustness of the solution. The closed-form solution is
\begin{equation}
x_{\lambda} = \left(\mathcal{M}^{T}\mathcal{M} + \lambda^{2} L^{T}L\right)^{-1} \mathcal{M}^{T}y,
\end{equation}
which reduces to the ordinary least-squares estimate in the limit $\lambda \to 0$ but remains stable for finite $\lambda$ even when $M$ is ill-conditioned.

A principled choice of $\lambda$ is critical. Too small a value results in noise amplification, while too large a value biases the solution toward the null space of $L$. We adopt the generalized cross-validation (GCV) criterion, which selects $\lambda$ by minimizing the predictive mean-square error across all possible leave-one-out partitions of the data. The GCV functional is defined as, \cite{hansen1998, hansen2010},
\begin{equation}
\mathrm{GCV}(\lambda) = 
\frac{\| (I - A_{\lambda})y \|_{2}^{2}}{\left[\mathrm{trace}(I - A_{\lambda})\right]^{2}},
\end{equation}
where $A_{\lambda} = \mathcal{M} \left(\mathcal{M}^{T}\mathcal{M} + \lambda^{2} I\right)^{-1} \mathcal{M}^{T}$ is the regularized influence matrix. The optimal parameter $\lambda^{\ast}$ is obtained as
\begin{equation}
\lambda^{\ast} = \arg\min_{\lambda > 0} \, \mathrm{GCV}(\lambda).
\end{equation}
This provides a statistically justified, data-driven method for balancing bias and variance without requiring explicit knowledge of the noise level.

\begin{figure*}[htb!]
\centering
\includegraphics[width=0.45\textwidth]{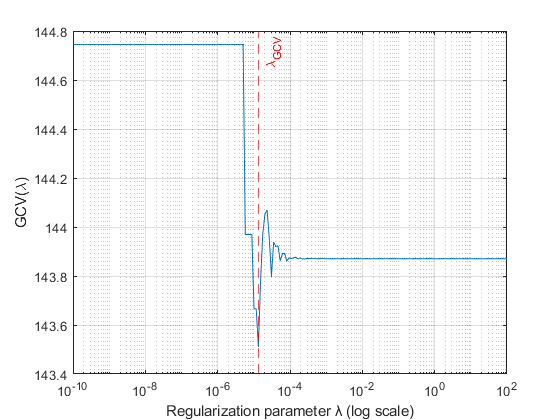}
\includegraphics[width=0.45\textwidth]{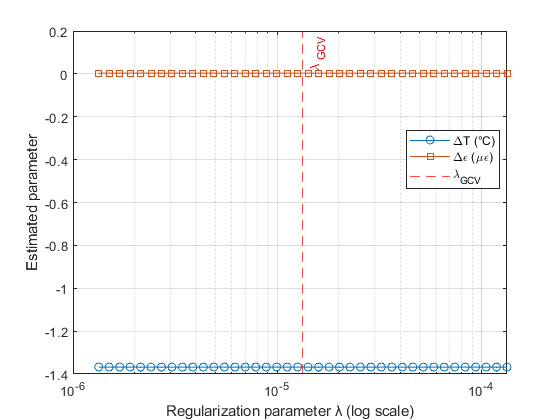}
\includegraphics[width=0.45\textwidth]{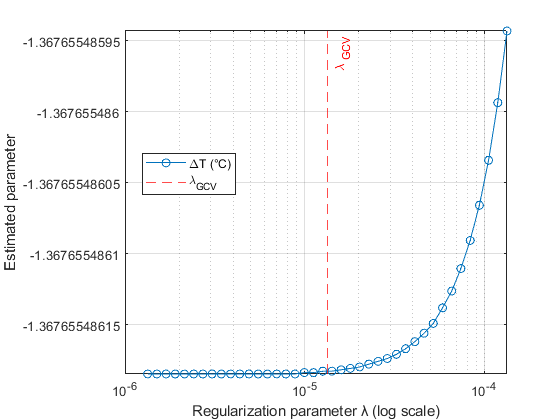} 
\includegraphics[width=0.45\textwidth]{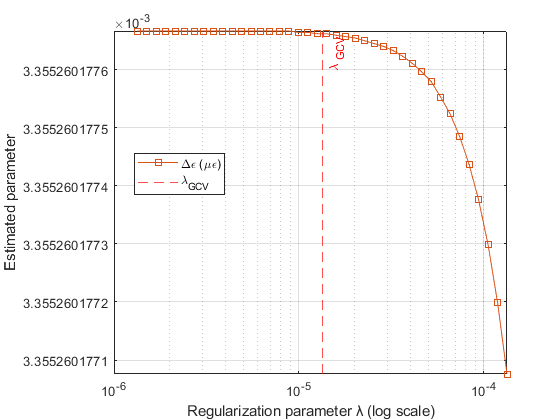}
\caption{\textbf{Selection of Tikhonov regularization parameter using GCV and stability verification.}
(a) GCV functional versus $\lambda$, with the optimal $\lambda^{\ast}$ marked at the minimum. (b) Estimated parameters $\Delta T$ and $\Delta \epsilon$ versus $\lambda$, showing extended plateaus. 
(c) Zoomed stability curve for $\Delta T$, demonstrating negligible variation ($\sim 10^{-8}\,\%$). 
(d) Zoomed stability curve for $\Delta \epsilon$, confirming invariance within $\sim 10^{-5}\,\%$. }
\label{fig:GCV}
\end{figure*}

The minimum of the GCV curve (Fig.~\ref{fig:GCV}(a)) yields $\lambda^{\ast}$, marked as the vertical dashed line. 
To further validate the robustness of the inversion, we compute stability curves of the recovered parameters as functions of $\lambda$. Specifically, we plot $\Delta T(\lambda)$ and $\Delta \epsilon(\lambda)$ obtained from Eq.~(18) over a broad range of $\lambda$. As shown in Figs.~\ref{fig:GCV}(b-d), both parameters exhibit extended plateaus around $\lambda^{\ast}$, remaining essentially invariant within $\pm 1$ decade. 

The existence of a plateau around $\lambda^{\ast}$, where the recovered parameters remain essentially invariant within $\pm 1$ decade of $\lambda^{\ast}$, is a strong indication of solution stability and immunity to overfitting. In our case, the variation was found to be vanishingly small (on the order of $10^{-8}\%$ for $\Delta T$ and $10^{-5}\%$ for $\Delta \epsilon$), confirming that the inversion results are not artifacts of arbitrary parameter tuning but reflect physically meaningful robustness.

\subsection{Simultaneous Measurement using Power-Based FBG Sensor}
After ensuring proper matrix conditioning and incorporating regularization to maintain numerical stability, the matrix-based framework was implemented for simultaneous estimation of multiple strain components and temperature variations. In Figs.~\ref{Error}(a-c) we present a comparison between the predicted and actual strain values under the influence of thermal noise in the 0-1 $^\circ$C range. Across all three strain parameters, the predicted data align remarkably well with the applied strains showing minimal deviation. The results closely trace the ideal diagonal line, confirming the robustness and accuracy of the method even in the presence of temperature-induced disturbances.

In Fig.~\ref{Error}(d-f), we have plotted the similarity figures showing the error (colorbar) between the simultaneously varying external perturbations applied to the sensor, and their estimated values obtained using the method discussed above (\begin{math} error (\mathcal{O}) = \Delta \chi_{perturbation} - \Delta \chi_{estimated} \end{math}) using Eq.~\ref{simultaneous}; the two simultaneously varying perturbation parameters have been taken as the environmental temperature, $T$, and axial strain $\epsilon$ (Fig. 4(a)), radial strain $\epsilon_r$ (Fig. 4(b)), and hydrostatic strain $\epsilon_H$ (Fig. 4(c)), respectively. 
As can be observed in this figure, a narrow range of temperature and strain has been examined, since within this interval the power response spans its full dynamic range, varying from near-zero to values approaching unity.
We observe that the error associated with the estimated perturbations for power based FBG sensor, as obtained using the above discussed inverse-problem approach, is of the order of $\mathcal{O} \sim$ $10^{-4}$ for axial strain, $10^{-6}$ for radial strain and $10^{-7}$ for hydrostatic strains, making it a powerful tool to estimate simultaneously varying perturbation parameters by tracking multiple resonances.

\subsection{Simultaneous Measurement using Wavelength-Based FBG Sensor}
In a wavelength-based FBG sensor, the sensor monitors the shift in the wavelength at the exceptional points, when the FBG experiences changes in a physical parameter such as strain or temperature. Figs.~\ref{Error1}(a-c) present a comparison between the predicted and actual strain values under the influence of thermal noise in the 0-20 $^\circ$C range. Across all three strain levels, the predicted data again align well with the applied strains, showing excellent consistency and minimal deviation. Similar to the power based sensor, in Fig.~\ref{Error1}(d-f) we have plotted the similarity figures showing the error (colorbar) between the simultaneously varying external perturbations applied to the sensor, and their estimated values obtained using the inverse-problem approach discussed above. Since the spectral interrogation technique is not limited by the power transitioning between maxima and minima (and is rather based on tracing the spectral evolution of EPs), a wider range of perturbations has been selected in this study.  
We observe that the error associated with the estimated perturbations for the wavelength-based FBG sensor is of the order of $\mathcal{O} \sim$ $10^{-3}$ for axial strain (Fig.~\ref{Error1}(d)) and radial strain (Fig.~\ref{Error1}(e)), and $\mathcal{O} \sim 10^{-4}$ for hydrostatic strains (Fig.~\ref{Error1}(f)), demonstrating the high sensitivity and reliability of this approach in precisely resolving simultaneous variations in two perturbation parameters. 
We also observe some hot-spots in Fig.~\ref{Error1}(d), which primarily occur when the two perturbations cause nearly collinear wavelength shifts, reducing the determinant of $\mathcal{M}$ and amplifying inversion errors.

For practical applications, when environmental noises couple into multiple parameters simultaneously, a systematic mitigation strategy can be adopted. One effective approach is to incorporate a reference FBG that serves as a baseline for tracking environmental variations. In addition, applying a digital low-pass filter can help suppress high-frequency fluctuations, thereby improving signal fidelity. Alternatively, calibration maps may be constructed to disentangle deterministic perturbations from stochastic noise, ensuring that the true signal of interest is preserved. Collectively, these methods can provide a robust framework for minimizing environmental interference and enhancing measurement accuracy.

\section{Conclusion}
In this work, we have demonstrated a novel approach for enhancing the sensitivity of Fiber Bragg Grating (FBG) sensors by leveraging the unique properties of exceptional points (EPs) in non-Hermitian photonic systems. By carefully engineering the grating parameters to position the spectral edges in proximity to EP conditions, we achieved substantial sensitivity enhancements across multiple perturbation parameters. The proposed sensors exhibited sensitivities of
$\sim9.227 dBm/^{\circ}C$, 0.4475 dBm/$\mu\epsilon$, 2.856 dBm/$\mu\epsilon$, and 0.182335 dBm/$\mu\epsilon$, which corresponds to improvements of approximately an order of magnitude, when compared to conventional FBG sensors.
In addition, we have also addressed the challenge of cross-sensitivity between temperature and various strains by implementing a sensitivity matrix-based interrogation scheme, enabling simultaneous and accurate discrimination of multiple perturbation effects. This multi-parameter sensing capability, combined with the enhanced sensitivity derived from EP physics, establishes EP-engineered FBGs as a powerful and versatile sensing platform.
The results presented here open new avenues for developing next-generation high-precision sensors with broad applicability across diverse fields, including telecommunications, aerospace structural monitoring, biomedical diagnostics, and environmental sensing. For field deployment, environmental robustness is critical. Coating damage, humidity ingress, or mechanical fatigue can shift EP alignment and degrade sensitivity. Standard protective strategies, including hermetic sealing, UV-stable polymer overcoats, and steel housing etc., effectively mitigate these risks. With such packaging, EP-enhanced FBGs can perform reliably in outdoor, aerospace, and underwater conditions. Future work may further explore dynamic tuning of EP conditions, integration with photonic integrated circuits, and the extension of this framework to complex multi-mode or hybrid sensing platforms.

\section*{Acknowledgment}
The authors gratefully acknowledge the computational resources provided by IIT Delhi.

\section*{Disclosures}
The authors declare no conflicts of interest.

\section*{Funding}
The authors declare that no funding was obtained for the preparation of this manuscript.

\ifCLASSOPTIONcaptionsoff
  \newpage
\fi

% trigger a \newpage just before the given reference
% number - used to balance the columns on the last page
% adjust value as needed - may need to be readjusted if
% the document is modified later
%\IEEEtriggeratref{8}
% The "triggered" command can be changed if desired:
%\IEEEtriggercmd{\enlargethispage{-5in}}

% references section

% can use a bibliography generated by BibTeX as a .bbl file
% BibTeX documentation can be easily obtained at:
% http://mirror.ctan.org/biblio/bibtex/contrib/doc/
% The IEEEtran BibTeX style support page is at:
% http://www.michaelshell.org/tex/ieeetran/bibtex/
%\bibliographystyle{IEEEtran}
% argument is your BibTeX string definitions and bibliography database(s)
%\bibliography{IEEEabrv,../bib/paper}
%
% <OR> manually copy in the resultant .bbl file
% set second argument of \begin to the number of references
% (used to reserve space for the reference number labels box)

\bibliographystyle{ieeetran}
\bibliography{Biblio_IEEE}

% biography section
% 
% If you have an EPS/PDF photo (graphicx package needed) extra braces are
% needed around the contents of the optional argument to biography to prevent
% the LaTeX parser from getting confused when it sees the complicated
% \includegraphics command within an optional argument. (You could create
% your own custom macro containing the \includegraphics command to make things
% simpler here.)
%\begin{IEEEbiography}[{\includegraphics[width=1in,height=1.25in,clip,keepaspectratio]{mshell}}]{Michael Shell}
% or if you just want to reserve a space for a photo:

\begin{IEEEbiography}{Neha Ahlawat}
...
\end{IEEEbiography}

\begin{IEEEbiography}{Saurabh Mani Tripathi}
...
\end{IEEEbiography}

% You can push biographies down or up by placing
% a \vfill before or after them. The appropriate
% use of \vfill depends on what kind of text is
% on the last page and whether or not the columns
% are being equalized.

%\vfill

% Can be used to pull up biographies so that the bottom of the last one
% is flush with the other column.
%\enlargethispage{-5in}

% that's all folks
\end{document}